\documentclass[a4paper]{jpconf}
\usepackage[utf8]{inputenc}
\usepackage{graphicx}
\usepackage{amsmath}
\usepackage{amssymb}
\def\be{\begin{equation}}
\def\ee{\end{equation}}
\def\ba{\begin{eqnarray}}
\def\ea{\end{eqnarray}}

\def\bi{\begin{itemize}}
\def\ei{\end{itemize}}

\begin{document}
%
%\linenumbers
%
\title{Ground-based Gamma-Ray Astronomy: an Introduction}

\author{Giuseppe Di Sciascio}

\address{INFN - Roma Tor Vergata}

\ead{giuseppe.disciascio@roma2.infn.it}

\begin{abstract}
During the last two decades Gamma-Ray Astronomy has emerged as a powerful tool to study cosmic ray physics. In fact, photons are not deviated by galactic or extragalactic magnetic fields so their directions bring the information of the production sites and are easier to detect than neutrinos. Thus the search for $\gamma$ primarily address in the framework of the search of cosmic ray sources and to the investigation of the phenomena in the acceleration sites.
This note is not a place for a review of ground-based gamma-ray astronomy.
We will introduce the experimental techniques used to detect photons from ground in the overwhelming background of CRs and briefly describe the experiments currently in data taking or under installation.
\end{abstract}
\section{Introduction}
\label{intro}

Understanding cosmic ray origin and transport through the interstellar medium is a fundamental problem which has a major impact on models of the structure and nature of the universe.
During the last years Gamma-Ray Astronomy and Neutrino-Astronomy have emerged as  powerful tools to study cosmic ray features. In fact these neutral particles are not deviated by galactic or extragalactic magnetic fields so their directions bring the information of the production sites. Thus the search for $\gamma$ or $\nu$ primarily address in the framework of the search of cosmic ray sources and to the investigation of the phenomena in the acceleration sites. Due to the extremely small cross sections for weak interactions ($\lambda \sim \frac{10^{12} }{E}$ g/cm$^2$), neutrinos
are able to leave compact sources, providing the observer with information from inside or the surroundings of supernovae, active galaxies or other cosmic systems. The challenge is to find neutrino induced muons among those produced by the primary cosmic rays in the atmosphere. 

\begin{figure}[ht!]
\begin{minipage}[t]{.47\linewidth}
  \begin{center}
\includegraphics[width=0.9\textwidth]{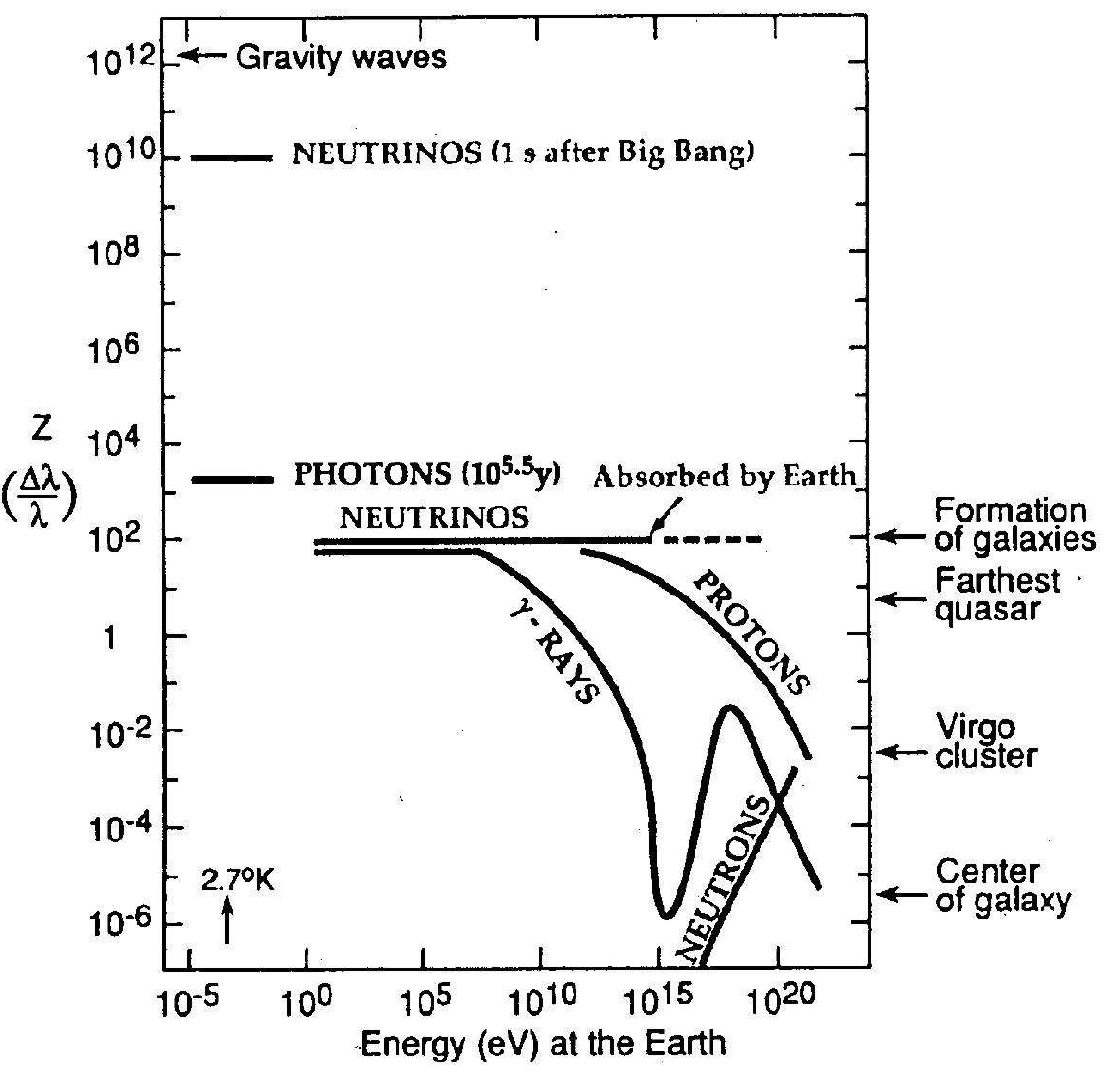}
  \end{center}
  \caption{The horizon measured in red-shift as a function of the energy at the earth of different kind of cosmic radiations.}
  \label{fig:cocconi}
\end{minipage}\hfill
\begin{minipage}[t]{.47\linewidth}
  \begin{center}
\includegraphics[width=0.9\textwidth]{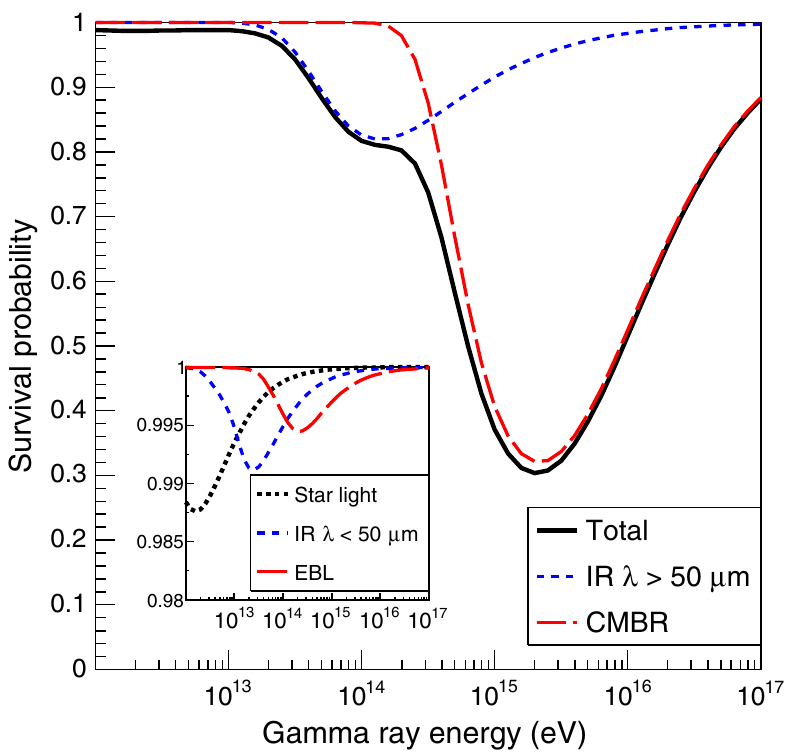}
  \end{center}
  \caption{Survival probability of gamma rays for a trajectory from the Galactic Center to the Sun, plotted as a function of the gamma ray energy. The contributions of different radiation fields are shown. The inset shows the contributions of starlight, infrared radiation with wavelength $\lambda< 50 \mu$m and EBL. Plot taken from \cite{vernetto-lipari}}
  \label{fig:gamma-absorption}
\end{minipage}\hfill
\end{figure}

Even if the horizon of a high energy photon is not extended, due to $\gamma - \gamma$ interaction, as the neutrino one (see Fig.~ref{fig:cocconi}), TeV $\gamma-$rays from many sources have been definitively observed providing evidence of the existence of energetic accelerating mechanisms.
Figure \ref{fig:gamma-absorption} shows the survival probability for gamma rays arriving to the Sun from a source in the Galactic center as a function of the gamma ray energy. In the figure the probability is shown together with the contributions from the different components of the radiation field. Most of the absorption is due to the cosmic microwave background radiation (CMBR) and to the thermal emission from the dust (with wavelength $\lambda\gtrsim$ 50 $\mu$m). The other components of the interstellar radiation field give smaller contributions that are visible in the inset of the figure. The survival probability has a deeper minimum P$_{surv}\approx$0.30 for E$_{\gamma}\sim$2.2 PeV that is due to the CMBR, and a second minimum at E$_{\gamma}\sim$ 50 TeV \cite{vernetto-lipari}.
%
%%%%%%%%%%%%%%%%%%%%%%%%%%%%%%%%%%%%%%%%%%%%%%%%%%%%%%%%%%%
%%%%%%%%%%%%%%%%%%%%%%%%%%%%%%%%%%%%%%%%%%%%%%%%%%%%%%%%%%%
%

The detection of high energy $\gamma-$rays is the final step to use the whole electromagnetic spectrum for investigating the most energetic processes and phenomena in the Universe.
Gamma-rays come from a variety of astronomical sources which accelerate charged cosmic rays (SuperNovae (SN), neutron stars, quasars, Active Galactic Nuclei (AGN)). Therefore, the detection of very high-energy (VHE) and ultrahigh-energy (UHE) $\gamma-$ray signals from celestial point sources give us a clue for understanding the acceleration of particles to ultrahigh-energies.

Therefore, an important motivation for high energy $\gamma-$ray astronomy is the study of the phenomenology of these sources. Since non-thermal spectra are a common feature of the observed $\gamma$ sources, detection of TeV sources can provide constraints on the models of acceleration and radiation processes at extreme conditions. The discovery of new types of sources is an intriguing perspective, in particular the discovery of sources of CRs where hadronic mechanisms are at work.

The riddle of the origin of CRs is unsolved since more than one century. The identification of the galactic sources able to accelerate particles beyond PeV (=10$^{15}$ eV) energies, the so-called \emph{'PeVatrons'}, is certainly one of the main open problems of high energy astrophysics. In fact, even there is no doubt that galactic CR are accelerated in SuperNova Remnants (SNRs), the capability of SNRs to accelerate CRs up to the \emph{'knee'} of the all-particle energy spectrum ($\sim$3$\times$10$^{15}$ eV) and above is still under debate.

Recently AGILE and Fermi observed GeV photons from two young SNRs (W44 and IC443) showing the typical spectrum feature around 1 GeV (the so-called \emph{'$\pi^0$ bump'}, due to the decay of $\pi^0\to\gamma\gamma$) related to hadronic interactions \cite{pizero-a,pizero-f}. 
This important measurement, however, does not demonstrate the capability of SNRs to produce the power needed to maintain the galactic CR population and to accelerate CRs up to the knee, at least. 
In fact, unlike neutrinos that are produced only in hadronic interactions, the question whether $\gamma$-rays are produced by the decay of $\pi^0$ from protons or nuclei interactions (\emph{'hadronic'} mechanism), or by a population of relativistic electrons via Inverse Compton scattering or bremsstrahlung (\emph{'leptonic'} mechanism), still needs a conclusive answer.
%
%%%%%%%%%%%%%%%%%%%%%%%%%%%%%%%%%%%%%%%%%%%%%%%%%%%%%%%%%%%%%%%%%%%%
\begin{figure}[ht!]
\begin{center}
\includegraphics[width=0.7\textwidth]{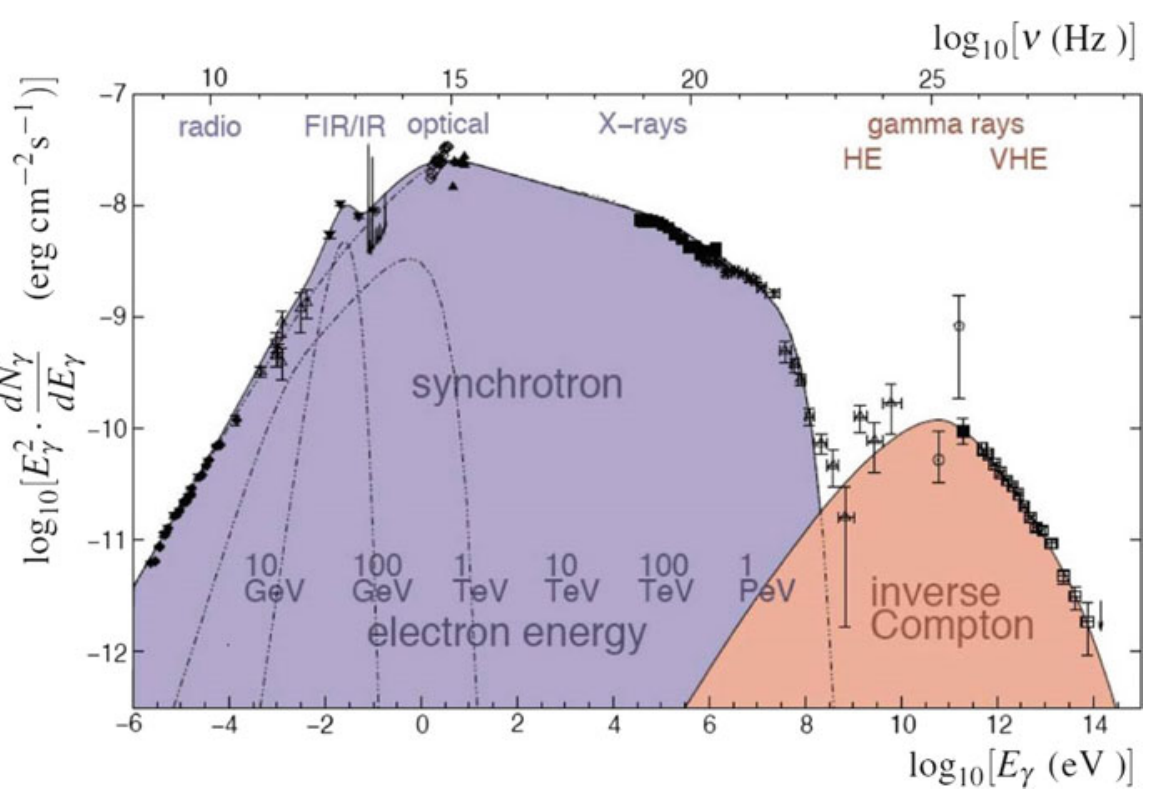}
\end{center}
\caption{Multiwavelength observations of the Crab nebula. The synchrotron emission is given by the superposition of the contributions of electrons with different energies. The synchrotron spectrum provides the target photons for the inverse Compton (IC) process \cite{spurio}. }
\label{fig:crab-sed}
\end{figure}
%%%%%%%%%%%%%%%%%%%%%%%%%%%%%%%%%%%%%%%%%%%%%%%%%%%%%%%%%%%%%%%%%%%%
%

High energy photons can be produced by different leptonic processes: bremsstrahlung, synchrotron radiation, inverse compton scattering.
Consider a population of relativistic electrons in a magnetized region. They will produce synchrotron radiation, and therefore they will fill the region with photons. These synchrotron photons will have some probability to interact again with the electrons, by the Inverse Compton process. 
Since the electron "work twice" (first making synchrotron radiation, then scattering it at higher energies) this particular kind of process is called \emph{Synchrotron Self-Compton}, or SSC for short. An example of a typical SSC spectrum is shown in Fig. \ref{fig:crab-sed}.
The \emph{one-zone model} assumes that non-thermal radiations are produced in a single, homogeneous and spherical region in the jet. 
The emission region moves relativistically toward us, and consequently the intrinsic radiation is strongly amplified due to the Doppler boosting. 
Three parameters are needed to characterise the emission region: the comoving magnetic field, the Doppler factor and the comoving radius of the emission region.
With one-zone SSC models we can describe practically all high energy gamma-ray emissions observed from Galactic and extra-galactic sources.

Discrimination between different models is very challenging.
In a hadronic interaction the secondary photons have an energy a factor of 10 lower than the primary proton. 
Therefore, the quest for CR sources able to accelerate particles in the knee energy region requires to survey the $\gamma$-ray sky above 100 TeV. In addition, the Inverse Compton scattering at these energies is strongly suppressed by the Klein-Nishina effect. Therefore, the observation of a $\gamma$-ray power law spectrum extending up to the 100 TeV range would be a strong indication of the hadronic nature of the emission.

To open the 100 TeV range to observations a detector with a very large effective area, wide field of view (FoV), operating with high duty-cycle, is required.
The most sensitive experimental technique for the observation of $\gamma$-rays at these energies and above is the detection of EAS via large ground-based arrays.
The muon content of photon-induced showers is very low, therefore these events can be discriminated from the large background of CRs via a simultaneous detection of muons that originate in the muon-rich CR showers. 
In addition, the large FoV and high duty cycle of EAS arrays make this observational technique particularly suited to perform unbiased all-sky surveys, and not simply observations of limited regions of the Galactic plane.

Gamma-ray astronomy could also give new hints to cosmology (through the measurement of the infrared background or studying primordial black holes) and particle physics beyond the Standard Model (evidence of supersymmetric particles or Dark Matter).

These grounds led to a large diffusion of the gamma-ray astronomy in the last decades. The number of present and planned experiments demonstrates the continuous growing interest of the scientific community in this field.
Satellite and Cerenkov detectors revealed a lot of $\gamma$-ray sources. However, to perform an \emph{"all sky"} monitoring looking at transient events, it is necessary a $\gamma$-rays detector with a wide FoV and high duty cycle. 

\begin{figure}[ht!]
\centering
\includegraphics[width=0.7\textwidth]{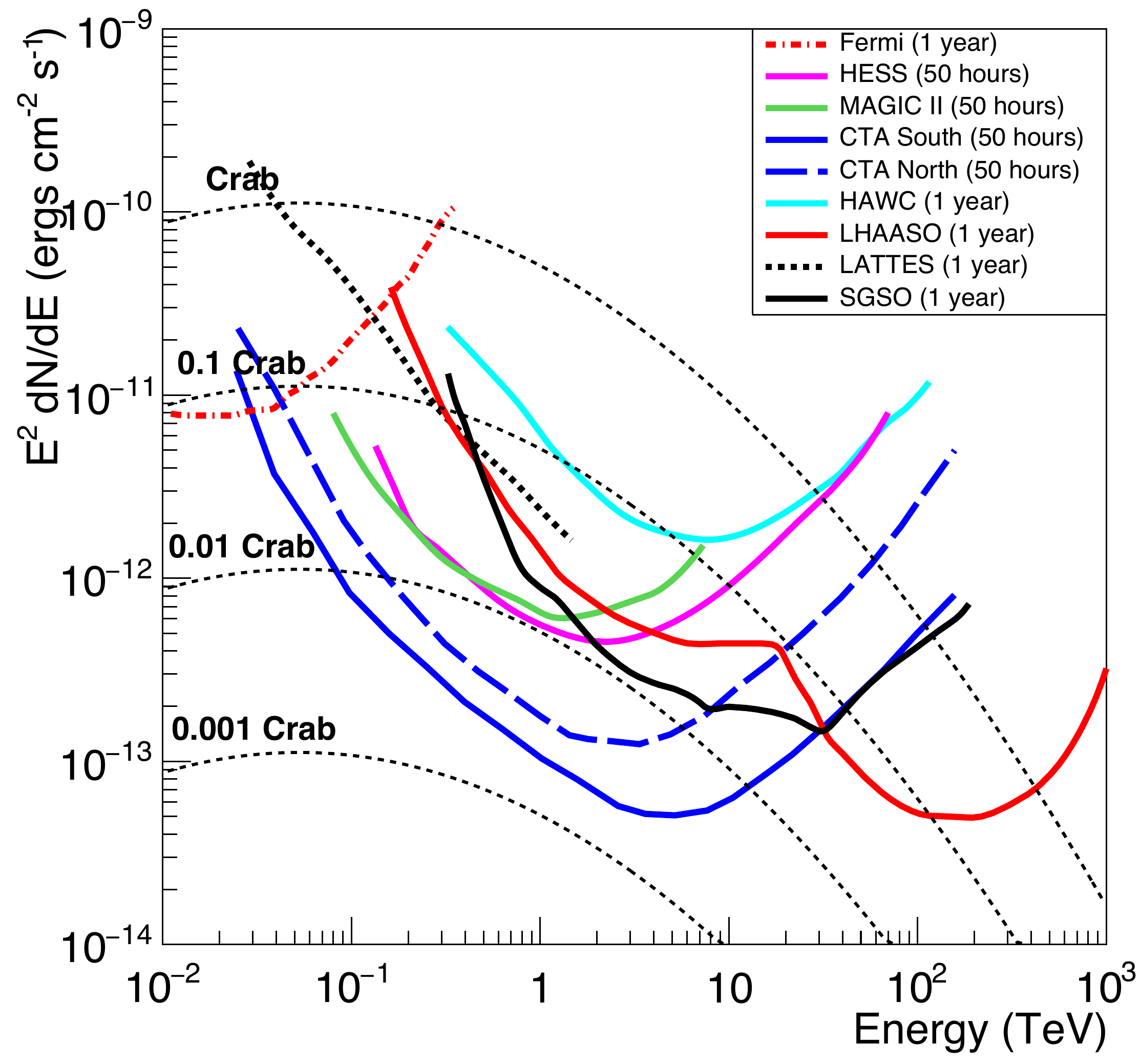}
\caption{Differential sensitivity (multiplied by E$^2$) to a Crab-like point gamma ray source of different experiments and projects. The best fit of the Crab Nebula data obtained by different detectors \cite{hhh15} is reported as a reference, extrapolated to 1 PeV.}
\label{fig:sensitivity}       % Give a unique label
\end{figure}

The differential sensitivity (multiplied by E$^2$) to a Crab-like point gamma ray source of different experiments and projects is shown in Fig. \ref{fig:sensitivity}.
The sensitivity of a detector is normally defined by the flux level $F_{min}$ (either photons cm$^{-2}$ s$^{-1}$ or percentage of Crab flux) for which a \emph{"5 $\sigma$ detection"} would be made (i.e. significance S = 5), and having at least 10 events, in 50 h of on-source observation (not counting any necessary additional off-source time) - or in the case of an unsteered detector such an Fermi-LAT or HAWC, in 1 year or 5
years of operation.
%
%%%%%%%%%%%%%%%%%%%%%%%%%%%%%%%%%%%%
%%%%%%%%%%%%%%%%%%%%%%%%%%%%%%%%%%%%
%

This is not a place for a review of ground-based gamma-ray astronomy (for which we recommend, for instance \cite{spurio,longair,aharonian-casanova,funk2015,aharonian2008,rieger2013,sinnis2009,hillas2013}).
We will introduce the experimental techniques used to detect photons from ground in the overwhelming background of CRs and briefly describe the experiments currently in data taking or under installation.

The paper is organized as follows:
The Section 2 we will devoted to a brief history of gamma-ray astronomy. The detection techniques will be summarized in Section 3 and the calculation of the sensitivity to a $\gamma$-ray source described in Section 4. 
In Sections 5 and 6 the Cherenkov and air shower techniques will be introduced and discussed.
New projects for an array in the Southern hemisphere will be briefly summarized in Section 7.

\section{Brief history of Gamma-Ray Astronomy}

At the beginning of the XVIIth century Galileo Galilei first used a telescope to observe the sky in a narrow spectrum of visible frequencies. Some centuries later, in the 1931-32, Jansky discovered radio cosmic emission. That was the beginning of the astronomy extended over all the electromagnetic spectrum.

In the summer of 1962, an X-ray detector on the Aerobee rocket was flown. The goal of this flight was originally to detect X-ray fluorescence of the Moon due to the Sun's wind. It failed to detect such emission, however, the detector serendipitously discovered extra solar X-rays from a source named Scorpius X-1 (the first X-ray source seen in the constellation Scorpio). In addition, it detected a diffuse X-ray background
(some of this has been resolved into sources and the rest is believed to be from unresolved active galaxies). 

In the 1961 the satellite Explorer XI first revealed $\gamma$-rays but the first significative results were obtained in the 1972-73 when the satellite SAS-2 took data for 7 months collecting about 8000 photons in the energy range 30 MeV $\div$ 5 GeV. In the period between the 1975 and the 1982 the satellite COS-B identified 25 galactic sources of photons at GeV energies.

In the late 60s Cocconi suggested the idea of exploring the TeV energy range using ground based detectors to observe $\gamma$-induced air showers \cite{cocconi}.
Indirect measurements at VHE and UHE were made from the 70s by means of Cerenkov telescopes. In the 1972, the Cerenkov telescope of the Crimean Astrophysical Observatory first revealed an excess of showers at TeV energies coming from the direction of the binary source Cygnus X-3. That signed the beginning of the VHE $\gamma$-ray astronomy \cite{weekes1988,hillas2013}.

In the early 80s the detection of UHE photons ($\sim$10$^{15}$ eV) from Cygnus X-3 has been claimed by Kiel and Haverah Park air shower arrays. However, a lot of dedicated ground based experiments (EAS-TOP, HEGRA, CASA-MIA) didn't detect any signal, and they provided only upper limits at a level of 1/100 of the previous detections. That seemed to mark the end of the high energy $\gamma$-ray astronomy \cite{weekes1988}.

\begin{figure}[ht!]
\begin{center}
\includegraphics[width=0.7\textwidth]{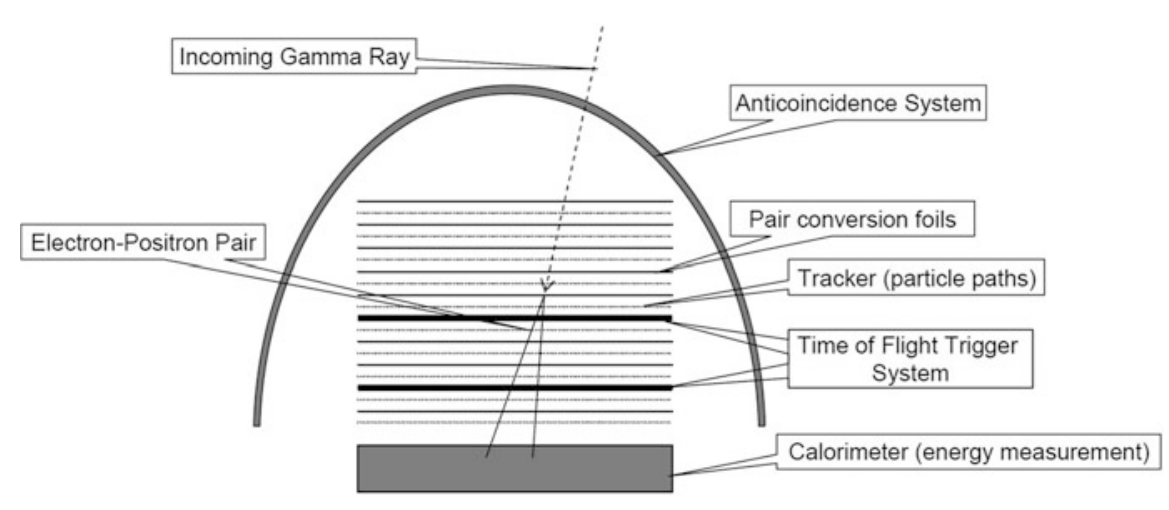}
\caption{Schematic diagram of a telescope which uses the conversion of a $\gamma$-ray into a e$^{\pm}$ pair (pair-conversion telescope), reproducing the features of the EGRET experiment. The Fermi-LAT is similar, with an improved tracker device \cite{spurio}.}
\end{center}
\label{fig:telescope-converter}
\end{figure}

The Compton Gamma Ray Observatory (CGRO) marked a turning point for $\gamma$-ray astronomy. 
A listing of the observations, along with other information about CGRO, can be found at the CGRO Science Support Center Web site http://cossc.gsfc.nasa.gov/ docs/cgro/index.html. 
The CGRO carried four instruments for $\gamma$-ray astronomy, each with its own energy range, detection technique, and scientific goals, covering energies from less than 15 keV to more than 30 GeV, over six orders of magnitude in the electromagnetic spectrum. The four instruments were:
\begin{itemize}
\item The Burst and Transient Source Experiment (BATSE). BATSE was the smallest of the CGRO instruments, consisting of eight modules located one on each corner of the spacecraft. Each unit included a large flat NaI(Tl) scintillator and a smaller thicker scintillator for spectral measurements, combined to cover an energy range from 15 keV to over 1 MeV.
\item Oriented Scintillation Spectrometer Experiment (OSSE). This used four large, collimated scintillator detectors to study $\gamma$-rays within the range from 60 keV to 10 MeV. OSSE mapped the 0.5 MeV line from positron annihilation and provided detailed measurements of many hard X-ray/soft $\gamma$-ray sources.
\item The Compton Telescope (COMPTEL), for the detection of medium energy $\gamma$-rays between 0.8 and 30 MeV, used a Compton scattering technique. Among its results, COMPTEL mapped the distribution of radioactive Aluminum-26 in the Galaxy, showing the locations of newly formed material.
\item The Energetic Gamma Ray Experiment Telescope (EGRET) was the high-energy instrument on CGRO, covering the energy range from 20 MeV to 30 GeV.
\end{itemize}
In the energy range above 10 MeV, the principal interaction process for $\gamma$-rays is pair production. Gamma rays cannot be reflected or refracted, and a high-energy $\gamma$-ray telescope detects e$^{\pm}$ with a precision converter-tracker section followed by a calorimeter.
The operational concept of EGRET, similar in most respects to the designs of other high-energy $\gamma$-ray telescopes, is shown in Fig. \ref{fig:telescope-converter}. The two key challenges for any such telescope are: (1) the identification of $\gamma$-ray interaction among the huge background of charged CRs; (2) the measurement of $\gamma$-ray energy, arrival time and arrival direction.
%
%%%%%%%%%%%%%%%%%%%%%%%%%%%%%%%%%%%%%%%%%%%%%%%%%%%%%%%%%%%%%%%%%%%
%%%%%%%%%%%%%%%%%%%%%%%%%%%%%%%%%%%%%%%%%%%%%%%%%%%%%%%%%%%%%%%%%%%%
%

In EGRET, these objectives were obtained as follows \cite{spurio}:
\begin{enumerate}
\item The charged particles were vetoed through the Anti-coincidence System (AS).The presence of a signal in the AS vetoed the tracking system electronics. A $\gamma$-ray candidate entered the detector without producing a signal in the anti-coincidence. The AS rejected nearly all unwanted signals produced by charged CRs. The AS consisted of a single dome of plastic scintillator, read out by 24 PMTs mounted around the bottom.
\item The $\gamma$-rays interacted in one of 28 thin sheets of high-Z material (tantalum), converting via pair production into an electron/positron pair.
\item A Tracker device, consisting of 36 wire grid spark chambers and interleaved with converter plates, was used to record the paths of the electron and positron, to reconstruct the conversion point and the arrival direction of the $\gamma$-ray. The spark chambers were gas detectors and their performance slightly deteriorated with time due to gas aging.
\item The electron and positron passed through two scintillator detectors. These fast-devices triggered the readout of the spark chambers and provided the time-of- flight (TOF) measurement (used to confirm the direction of the particles).
\item The electron and positron entered the Calorimeter, producing an electromagnetic shower. Using standard methods derived from particle physics, the energies of the particles were measured and, therefore, the energy of the parent $\gamma$-ray determined. The calorimeter was made of 36 NaI crystals read out by 16 PMTs.
\end{enumerate}
The EGRET angular resolution was $\sim 6^{\circ}$ at an energy of 100 MeV, and sources were localized with resolution of about 15 arcmin. The telescope has a field-of-view of about 0.5 sr. During its 9-years live time, EGRET detected over 1.5$\times$10$^6$ $\gamma$-rays, allowing it to build up the first picture of the entire high-energy $\gamma$-ray sky. 

\begin{figure}[ht!]
\begin{center}
\includegraphics[width=0.7\textwidth]{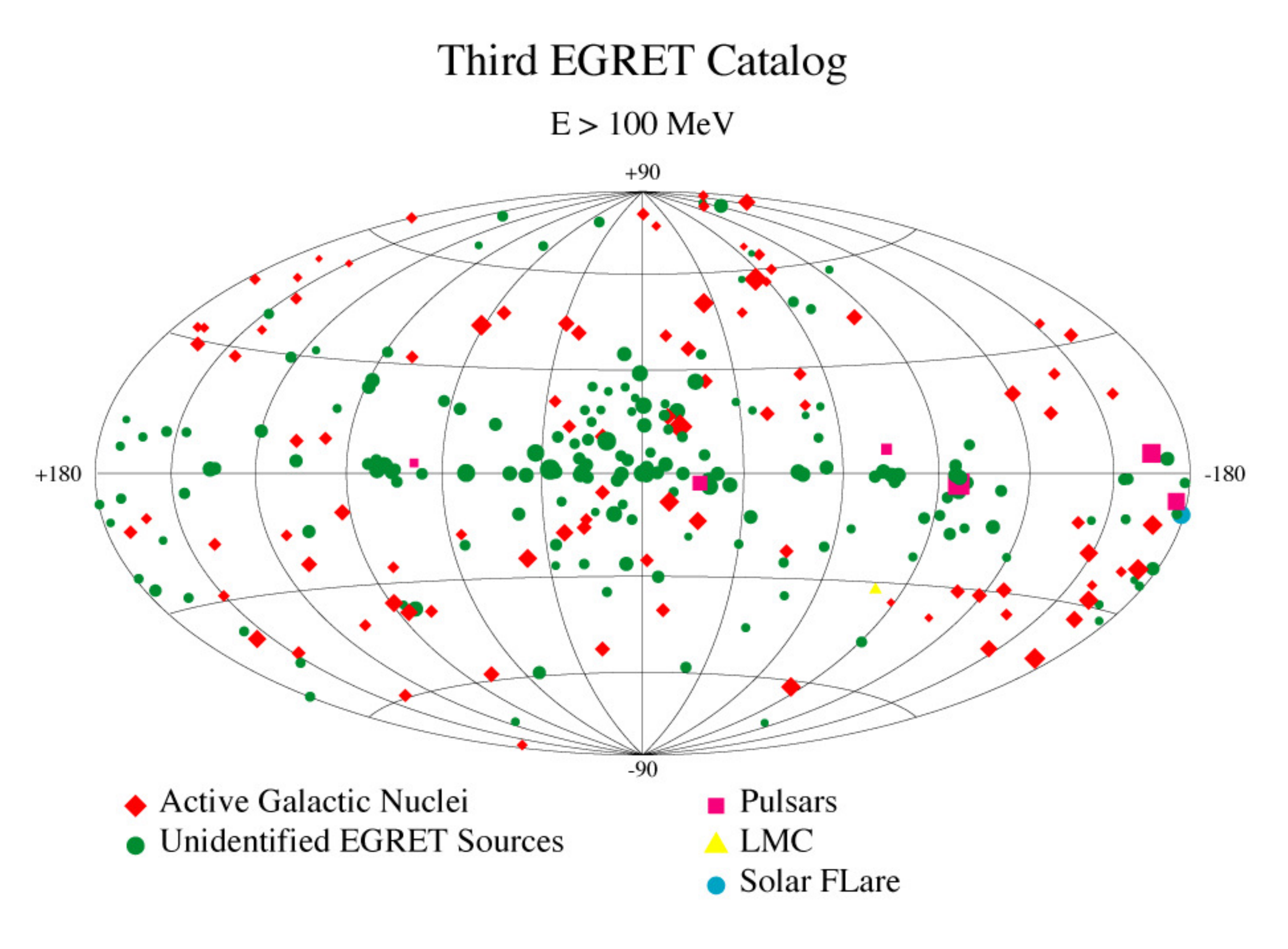}
\end{center}
\caption{Catalogue in galactic coordinates of the $\gamma$-ray point sources detected by EGRET above 100 MeV \cite{hartman1999}.}
\label{fig:c1_egret_map}
\end{figure}

The CGRO, launched into orbit in 1991 and operated successfully until it was de-orbited on June 4, 2000, thanks to the detector EGRET detected 271 sources with $\gamma$-rays at energies between 100 MeV and 10 GeV. Figure \ref{fig:c1_egret_map} shows a galactic map of the $\gamma$-ray point sources detected by EGRET as published in their third catalogue \cite{hartman1999} based on $\sim$4 years of observation (April 1991 - October 1995).
%
%%%%%%%%%%%%%%%%%%%%%%%%%%%%%%%%%%%%%%%%%%%%%%%%%%%%%%%%%%%%%%%%%%%
%%%%%%%%%%%%%%%%%%%%%%%%%%%%%%%%%%%%%%%%%%%%%%%%%%%%%%%%%%%%%%%%%%%%
%

The identified sources consist of 8 pulsars, 1 solar flare, 66 high-confidence blazars identifications, 27 possible blazars, 1 radio galaxy (Cen A) and the Large Magellanic Cloud (LMC). The remaining 170 sources, mostly located along the galactic plane, are not identified with known objects, since they have no observed counterparts at other wavelengths.
%
%%%%%%%%%%%%%%%%%%%%%%%%%%%%%%%%%%%%%%%%%%%%%%%%%%%%%%%%%%%%%%%%%%%
\begin{figure}[ht!]
\begin{center}
\includegraphics[width=0.7\textwidth]{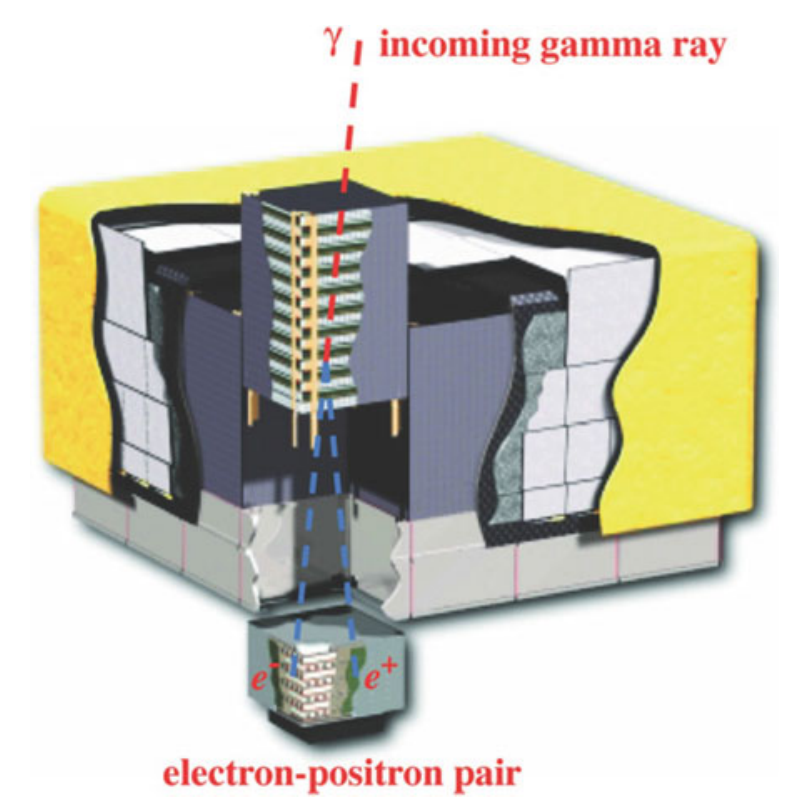}
\end{center}
\caption{Schematic diagram of the LAT. The telescope's dimensions are 1.8$\times$1.8$\times$0.72 m. The power required and the mass are
650 W and 2789 kg, respectively. The upper part is the tracker, the lower part the calorimeter and the surrounding region (in yellow) the veto \cite{spurio}.}
\label{fig:fermi}
\end{figure}
%%%%%%%%%%%%%%%%%%%%%%%%%%%%%%%%%%%%%%%%%%%%%%%%%%%%%%%%%%%%%%%%%%%%
%
%
%%%%%%%%%%%%%%%%%%%%%%%%%%%%%%%%%%%%%%%%%%%%%%%%%%%%%%%%%%%%%%%%%%%
\begin{figure}[ht!]
\begin{center}
\includegraphics[width=0.8\textwidth]{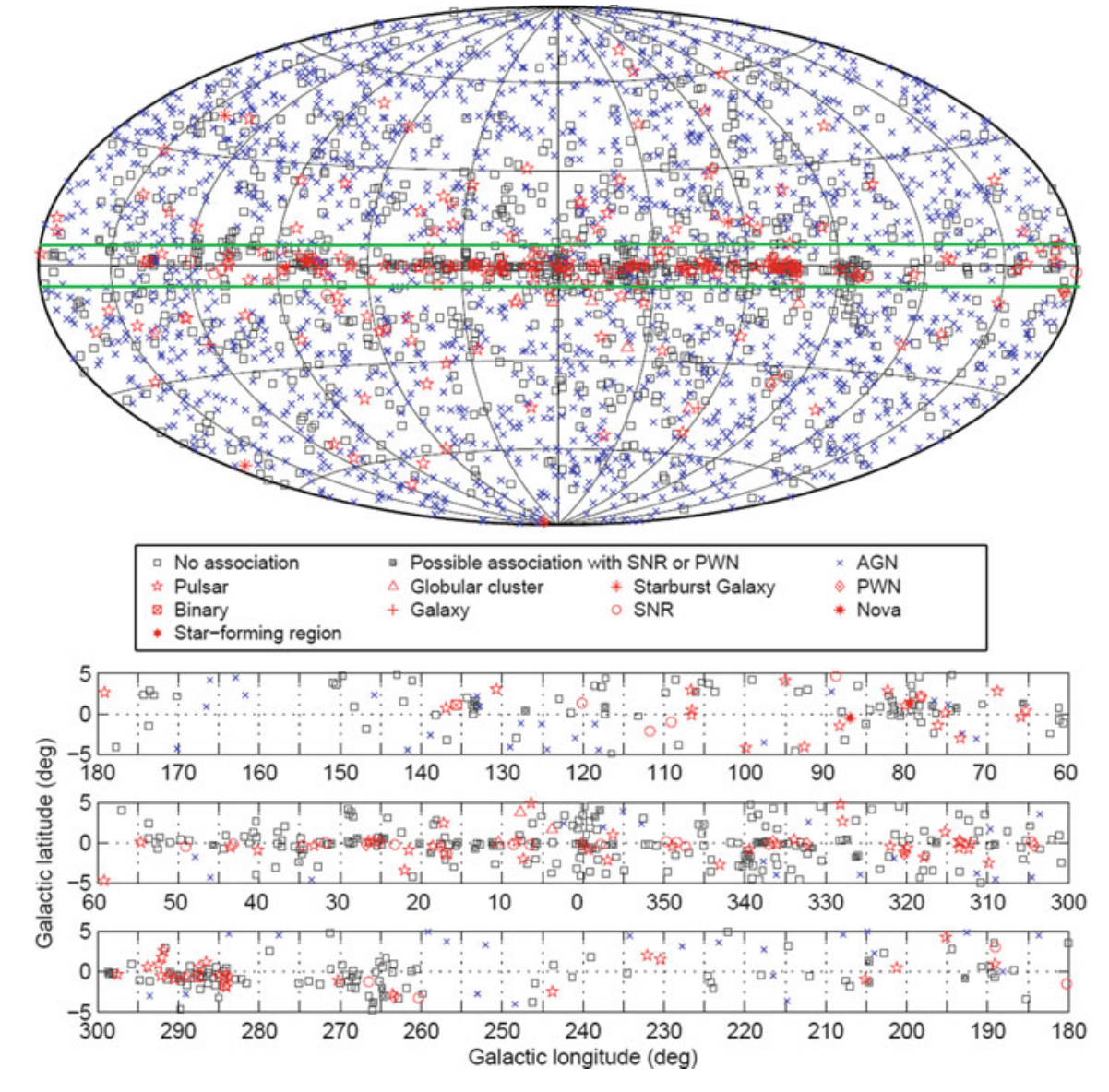}
\end{center}
\caption{Full sky map (top) and blow-up of the inner galactic region (bottom) showing sources by source class in the 3 FGL Fermi Catalog. Identified sources are shown with a red symbol, associated sources in blue. All AGN classes are plotted with the same symbol for simplicity \cite{spurio}.}
\label{fig:fermi-3fgl}
\end{figure}
%%%%%%%%%%%%%%%%%%%%%%%%%%%%%%%%%%%%%%%%%%%%%%%%%%%%%%%%%%%%%%%%%%%%
%

After EGRET, the situation regarding GeV $\gamma$-ray astronomy had a breakthrough in 2008 with the launch of the Fermi satellite \cite{fermi-lat}.
In fact, immediately after its launch, on June 11 2008, an overwhelming amount of data has significantly improved our knowledge of high energy astrophysics. The Large Area Telescope (LAT) is a telescope for $\gamma$-rays within the energy range from 20 MeV to more than 300 GeV, detected over the large background of energetic charged particles at the 565 km altitude orbit of the Fermi satellite. For each $\gamma$-ray, the LAT measures its arrival time, direction, and energy.

Fundamental in Fermi is the rejection capability to discriminate between electromagnetic and hadronic showers, based on the different event topology in the three subsystems (the tracker, the calorimeter and the anti-coincidence, as shown in Fig. \ref{fig:fermi}).
The LAT is a pair-conversion telescope with a precision converter-tracker section followed by a calorimeter. These two subsystems each consist of a 4$\times$4 array of 16 modules (see Fig. \ref{fig:fermi}). The large field-of-view results from the low aspect ratio (height/width) of the LAT made possible by the choice of particle tracking technology (i.e., silicon-strip detectors) that allowed for elimination of the time-of-flight triggering system used in EGRET.
The effective collecting area is $\sim$6500 cm$^2$ at 1 GeV with a wide field-of-view ($\sim$2 sr).

The Fermi collaboration is producing a large number of important scientific results that cannot be summarized in this note.
The third catalog of high-energy $\gamma$-ray sources (3FGL) detected by LAT \cite{Acero2015} includes 3033 sources  above 4 $\sigma$ significance within the 100 MeV - 300 GeV range (see Fig. \ref{fig:fermi-3fgl}). Among them, 238 sources are considered as having been identified based on angular extent or correlated variability (periodic or otherwise) observed at other wavelengths. For 1010 sources, no plausible counterparts at other wavelengths have been found. More than 1100 of the identified or associated sources are active galaxies of the blazar class. Pulsars represent the largest Galactic source class.
LAT collects about 150 million $\gamma$-rays per year (compared with the 1.5 million detected by EGRET in 9 years).
The minimum flux from a source that LAT can discriminate from the background corresponds to $\sim$3$\times$10$^{-12}$ erg cm$^{-2}$ s$^{-1}$ \cite{Acero2015}.

Fermi reported also important observations of Gamma-Ray Bursts (GRBs), extremely intense and relatively short bursts of gamma radiation that occur a few times per day in the detectable Universe. Their emission exceeds the gamma emission of any other source.
Roughly one GRB per week is detected with the Fermi-GBM between 8 keV - 30 MeV, and roughly one a month is detected with the Fermi-LAT, 20 MeV - 300 GeV. Several bursts have been detected by the Fermi-LAT at energies above 1 GeV (11 GRBs from August 2008 to January 2010), considerably improving our knowledge of high energy $\gamma$-ray emission. For further details and for a history of the discovery of GRBs we suggest the textbook by Spurio \cite{spurio}.

The Fermi satellite was anticipated by the smaller-scale telescope \emph{Astro-rivelatore Gamma a Immagini Leggero (AGILE)}. AGILE is a project of the Italian Space Agency (ASI) and was launched in April 2007. It is devoted to $\gamma$-ray observations within the 30 MeV-50 GeV energy range, with simultaneous hard X-ray imaging in the 18-60 keV band, and optimal timing capabilities for the study of transient phenomena. The AGILE instrument \cite{Tavani2008} consists of the Silicon Tracker, the X-ray detector SuperAGILE, the CsI(Tl) Mini-Calorimeter and an anti- coincidence system. The combination of these instruments forms the Gamma-Ray Imaging Detector (GRID). The very large field-of-view (2.5 sr) of the $\gamma$-ray imager coupled with the hard X-ray monitoring capability makes AGILE well suited to study galactic and extragalactic sources, as well as GRBs and other fast transients. AGILE reaches its optimal performance near 100 MeV with good imaging and sensitivity. Gamma-ray and hard X-ray sources can be monitored 14 times a day, and an extensive database has been obtained for a variety of sources.

At the end of 80s a new generation of Cerenkov telescopes enabled to establish that the Crab Nebula is a stationary and continuous source of photons with an energy from 0.5 TeV to 10 TeV \cite{hillas2013}.
Indeed in 1989 the Cerenkov telescope Whipple observed an excess of photons with energies $\ge$500 GeV coming from the direction of the Crab Nebula. After Whipple, other Cerenkov telescopes confirmed the stationary flux from the Crab, which is now considered the \emph{"standard candle"} for the northern hemisphere detectors. Only recently, $\gamma$-ray flares have been detected by AGILE and Fermi-LAT and pulsed emission from the Crab pulsar up to beyond 100 GeV observed by MAGIC and VERITAS. The origins of these variations are still under investigation.
%
%%%%%%%%%%%%%%%%%%%%%%%%%%%%%%%%%%%%%%%%%%%%%%%%%%%%%%%%%%%%%%%%%%%%
\begin{figure}[ht!]
\begin{center}
\includegraphics[width=0.7\textwidth]{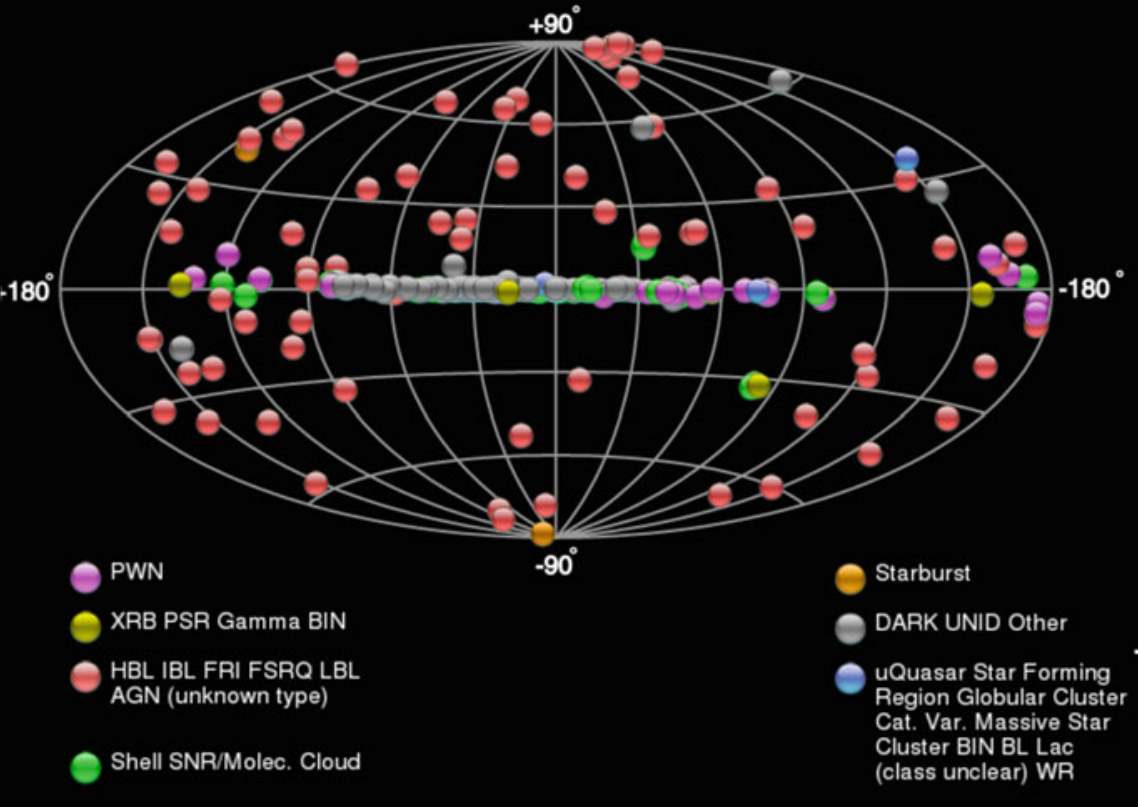}
\end{center}
\caption{Map of the 208 TeV sources (2018) retrieved from the online TeV catalog (http://tevcat. uchicago.edu/), which displays, with different color codes, the position in galactic coordinates of the various $\gamma$-ray sources detected from the ground. Most of the detections were performed by IACTs, which typically are sensitive to $\gamma$-rays well above 100 GeV.}
\label{fig:Tev_catalog}
\end{figure}
%%%%%%%%%%%%%%%%%%%%%%%%%%%%%%%%%%%%%%%%%%%%%%%%%%%%%%%%%%%%%%%%%%%%
%

Observations of the Crab are used for cross-calibration of ground-based detectors, to demonstrate the stability of a detector and to improve the data analysis methods. The Crab photon flux is considered a unit of measurement to evaluate the sensitivity of the apparatus. 
In the Figure \ref{fig:crab-sed} the Crab flux measured by different experiments at different energies is shown. The integral flux above 1 TeV corresponds to
\be
\Phi_{\gamma}^{Crab}(>1 TeV) = (2.1\pm 0.1)\times 10^{-11}\>\>\> {\rm photons\> cm^{-2} \> s^{-1}}
\ee
In 1999, the experiment Tibet AS-$\gamma$ provided the first observation of TeV photons with an air shower array.
In the following years Milagro and ARGO-YBJ demonstrated that shower arrays are able to detect gamma-ray sources and flaring emissions from extra-galactic sources. As will be discussed in Section \ref{sect:array}, they pushed the construction of a new generation of arrays, HAWC, LHAASO and HiSCORE.

However, the most successful instruments in the history of VHE $\gamma-$ray astronomy have been the Imaging Atmospheric Cherenkov Telescopes (IACTs), Section \ref{sect:iact}. In the last decade HESS, MAGIC and VERITAS experiments obtained a lot of important results in gamma-ray astronomy that cannot be summarized in this note (for further details see, for example, \cite{aharonian-casanova} and references therein).
 %
%%%%%%%%%%%%%%%%%%%%%%%%%%%%%%%%%%%%%%%%%%%%%%%%%%%%%%%%%%%%%%%%%%%%
\begin{figure}[ht!]
\begin{center}
\includegraphics[width=0.8\textwidth]{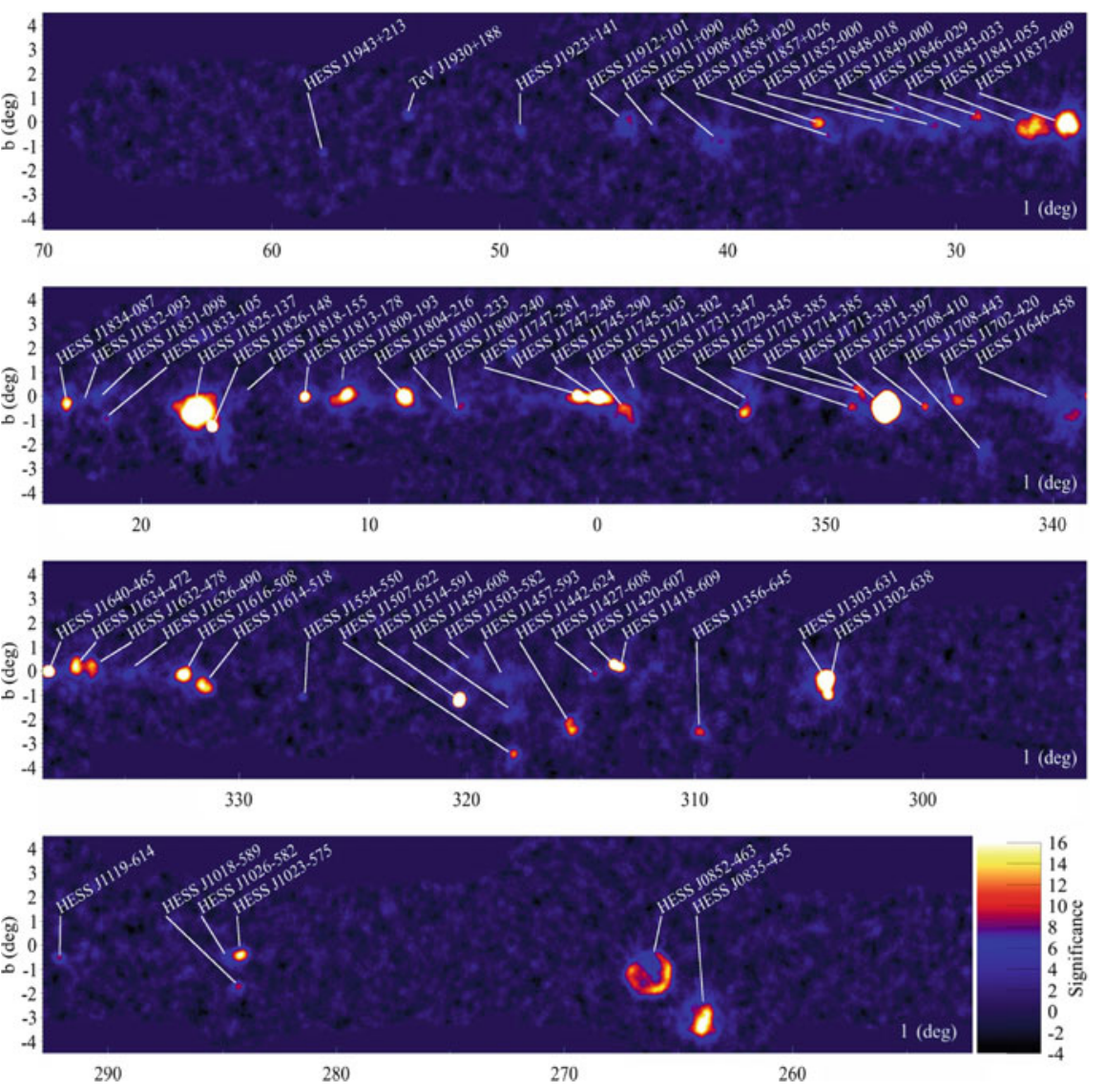}
\end{center}
\caption{Significance map of $\gamma$-ray objects in the galactic plane as observed by HESS \cite{hess-survey2018}.}
\label{fig:hess-survey}
\end{figure}
%%%%%%%%%%%%%%%%%%%%%%%%%%%%%%%%%%%%%%%%%%%%%%%%%%%%%%%%%%%%%%%%%%%%
%

The sky positions of the 208 known TeV-emitting sources (early 2018), mainly discovered by IACTs, are shown in Fig. \ref{fig:Tev_catalog}.
Most of the sources are associated with objects already known through different wavelength observations.
However, new classes of emitting sources have been discovered, the fraction of unassociated TeV sources is $\sim$25\%.

At TeV energies, the unidentified sources lie essentially on the galactic plane (only 4 out of 54 are far from the plane), as opposed to sources observed at the GeV energy range. This could be due to an observational bias. The wide field-of-view and the survey mode operation of the Fermi satellite provides LAT with a roughly (within a factor two) uniform exposure to the entire sky. 
The IACTs have very narrow effective fields-of-view and the galactic plane is the only large fraction of the sky that has been studied by HESS in detail. This dedicated survey of about 2700 h from 2004 to 2013 has covered the range in galactic longitude between [-85$^{\circ}$, +60$^{\circ}$] and [-3.5$^{\circ}$, +3.5$^{\circ}$] in latitude \cite{hess-survey2018}. The HESS angular resolution is $\approx$0.08$^{\circ}$ ($\approx$5 arcmin), the sensitivity $\leqslant$1.5\% Crab flux and the energy range 0.2 TeV $\to$ 100 TeV.
It has revealed more than fifty VHE $\gamma$-ray sources. 
The significance map of the discovered objects is shown in Fig. \ref{fig:hess-survey} \cite{hess-survey2018}. A large fraction (more than half) corresponds to Pulsar Wind Nebulae, located in close vicinity to young and energetic pulsars \cite{spurio}.

\section{Detection techniques}
\label{detectors}

The large energy range that can be investigated in gamma-ray astronomy ($\approx$MeV $\to$ PeV) implies a great variety of generation phenomena and require different detection techniques.
The experimental techniques that can be used in HE/VHE $\gamma-$ray astronomy are determined by the properties of the gamma radiation and by the background:
\begin{enumerate}
\item[(1)] The $\gamma$-ray flux is very small ($\le 10^{-3}$ with respect to the background of CRs detected in a $1^{\circ}$ angle around the direction of the source) and rapidly decreases when energy rises. All the known sources exhibit a typical differential spectrum in the form of a power-law:
\begin{equation}
\frac{dN}{dE} \propto E^{-\gamma}
\end{equation}
with $\gamma\sim$ 2 - 3. Small detectors onboard satellites permit observations of tenths GeV $\gamma$-rays. To reveal the low flux at higher energies it is necessary to build up ground-based detectors where is possible to have big collecting areas. 
\item[(2)] The Earth's atmosphere is opaque to $\gamma-$rays being about 28 radiation lengths thick at sea level. Therefore, $\gamma-$rays cannot be directly observed by ground-based detectors.
 \item The isotropic CR flux forms a formidable background exceeding by many orders of magnitude even the strongest steady photon flux. It
consists largely in protons and heavier nuclei.
\end{enumerate}
Given these limitations, a terrestrial observer must use different strategies to face the problem of the gamma detection.

\begin{figure}[ht!]
\begin{center}
\includegraphics[width=0.7\textwidth]{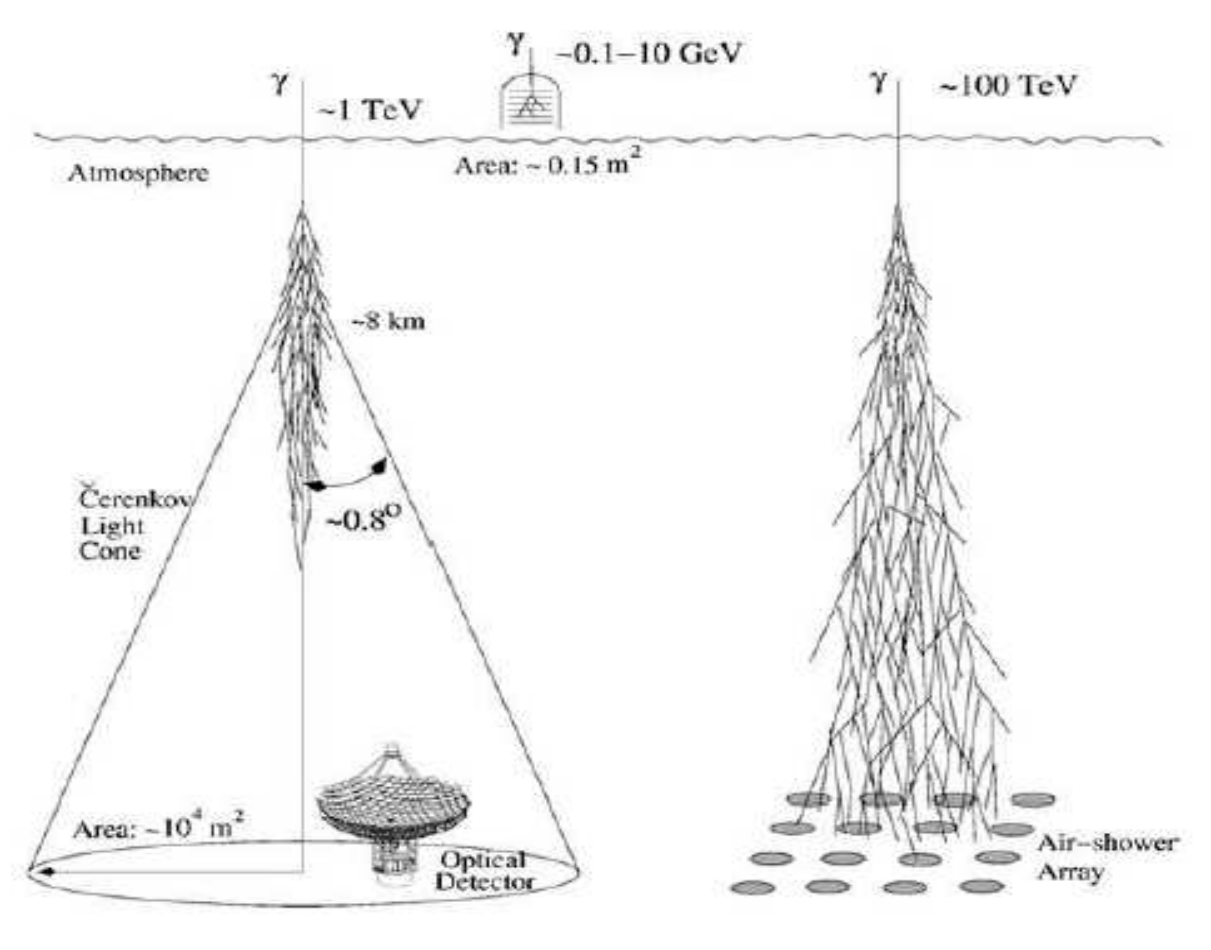}
\end{center}
\caption{Experimental techniques used for the detection of high energy $\gamma-$rays.}
\label{fig:chere-array}
\end{figure}

\begin{itemize}
\item Satellites:\\
Making use of detectors flown on satellites is the simplest way to avoid the problem of the atmosphere. As mentioned in Section 2, satellite detectors consist mainly of one or more converter layers in which HE photons produce a pair $e^{+}e^{-}$, a tracking detector in which the electron/positron is traced (used to reconstruct the direction of the incident photon) and a total absorption calorimeter that allows for an energy estimate (resolution $\le$ 20 \%). The problem of CR background is solved with a charged particle veto counter that efficiently rejects charged CRs. In the scheme shown in Figure \ref{fig:telescope-converter} the various elements can be recognized.\\
The strong limitation for this technique is the point (1): since the size of detectors is constrained by the weight that can be placed on satellites, their collection area is not large. Hence the rapid decreasing of the typical fluxes determines a maximum energy at which the collection area suffices for a statistical significant detection. 
\item Ground-based detection:\\
In the VHE range the low fluxes and the spectral slope of the typical sources require the use of very large detectors.
Therefore, observations must be done from the Earth's surface. 
The ground-based detection of VHE photons is indirect: nature, direction and energy of the primary particle have to be inferred from the measurable properties of the secondary particles (in the case of shower arrays) or of the Cherenkov flash (for Cherenkov telescopes).

Due to the opacity of the atmosphere to VHE photons, in fact, only the secondary effects of the atmospheric absorption can be detected. 
The mean free path of photons for pair production is almost the same as the electron radiation length, $X_0\sim$37 g/cm$^{-2}$ in air.
Gamma-rays interact electromagnetically, producing an electron/positron pair. These secondary particles yield a new generation of $\gamma$-rays through bremsstrahlung, starting the generation of an electromagnetic cascade. 

Any secondary charged particle in the shower produces Cherenkov light if its velocity exceeds the threshold $\beta = v/c>n$, where $n$ is the refraction index of atmosphere. 
The light is emitted at the Cherenkov angle $\theta$, with cos$\theta = 1/\beta n$. As the refraction index $n$ of the atmosphere changes with atmospheric depth, the Cherenkov angle increases from 0.66$^{\circ}$ at a height of 10 km to 0.74$^{\circ}$ at 8 km. This results in a rough focusing of light onto the ground into a ring-like region with radius of R$\sim$10 km$\times$ 0.012 rad = 120 m for a typical $\gamma$-ray shower. The number of Cherenkov photons emitted per unit length is $n_C\sim$0.1 photons cm$^{-1}$ at sea level. Multiplying $n_C$ by the number of particles at maximum ($N_{max}$) and by the path length of shower particles, the total number of Cherenkov photons turns out to be $N_C\sim 10^6$ for 1 TeV $\gamma$-rays. $N_C$ is proportional to $E_{\gamma}$.
\end{itemize}
Until now two experimental approaches have been used to detect $\gamma$ radiation from the ground (Fig. \ref{fig:chere-array}):
\begin{itemize}
\item Measurement of the Cherenkov light by means of telescopes.
\item Sampling of the charged particles of the air showers using arrays made by a large number of detectors scattered over wide areas.
\end{itemize}
%
%%%%%%%%%%%%%%%%%%%%%%%%%%%%%%%%%%%%%%%%%%%%%%%%%%%%%%%%%%%%%%%%%%
%%%%%%%%%%%%%%%%%%%%%%%%%%%%%%%%%%%%%%%%%%%%%%%%%%%%%%%%%%%%%%%%%%

\subsection{Background discrimination from ground}

The main drawback of ground-based measurement consists in the huge background of charged CRs.
This means that ground-based instruments detect a source as an excess of events from a certain direction over an overwhelming uniform background. For this reason their sensitivities are expressed in units of standard deviations of the CR background (for a discussion about sensitivity in $\gamma$-ray astronomy see next Section):
\begin{equation}
S = \frac{N_{\gamma}}{\sqrt{N_{bkg}}}
\label{sens1}
\end{equation}
To give an idea of the background level let us compare the flux of photons from a point source like the Crab Nebula with the isotropic flux of CRs within an angle of 1 deg around the direction of the source:
\be
\Phi_{\gamma}^{Crab}(>1\> {\rm TeV})\approx 2\cdot 10^{-11}\>\> {\rm photons\>\> cm^{-2} s^{-1}}
\ee
\be
\Phi_{CR}(>1 \>{\rm TeV})\Delta\Omega(\rm =1\> msr)\approx 2\cdot 10^{-8}\>\> {\rm nuclei\>\> cm^{-2} s^{-1}.}
\ee

As it can be seen, 
\be
\Phi_{\gamma}^{Crab}\approx 10^{-3}\cdot \Phi_{CR}
\ee

any $\gamma$-ray signal is completely overwhelmed by showers produced by ordinary charged CRs (mainly protons) spread evenly over the sky. 
And no veto with an anticoincidence shield, as in satellite experiments, is possible from ground.
In addition showers induced by photons are very similar to showers induced by charged nuclei.
Fortunately, some difference does exist. 
%
%%%%%%%%%%%%%%%%%%%%%%%%%%%%%%%%%%%%%%%%%%%%%%%%%%%%%%%%%%%%%%%%%%
\begin{figure}[ht!]
\begin{center}
\includegraphics[width=0.7\textwidth]{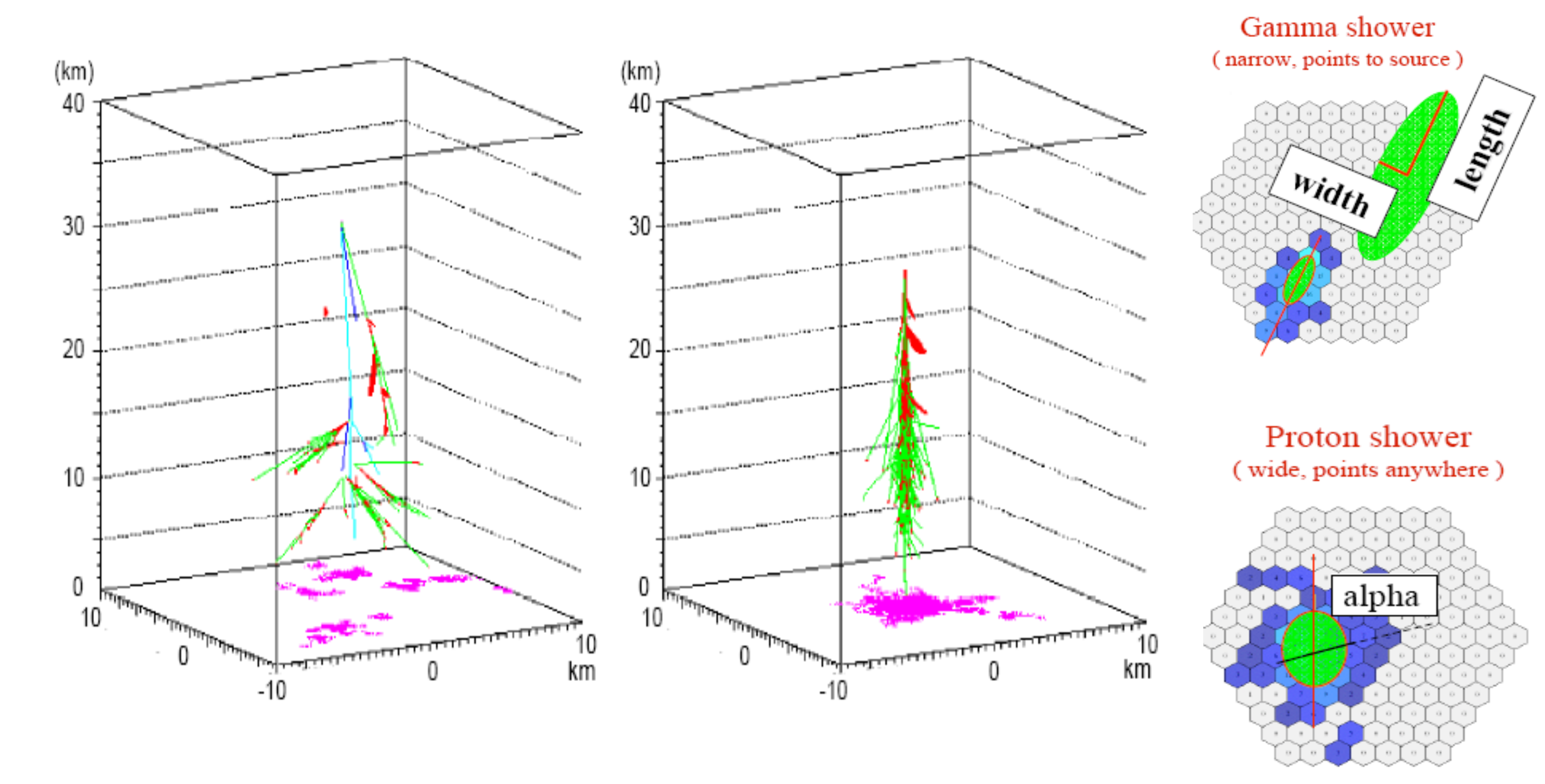}
\end{center}
\caption{Scheme of the lateral development of air showers induced by protons (left side) and photons (right side). The different lateral structures induce different images on the pixel camera of a IACT.}
\label{fig:cerenkov-scheme}
\end{figure}
%%%%%%%%%%%%%%%%%%%%%%%%%%%%%%%%%%%%%%%%%%%%%%%%%%%%%%%%%%%%%%%%%%
%

%
%%%%%%%%%%%%%%%%%%%%%%%%%%%%%%%%%%%%%%%%%%%%%%%%%%%%%%%%%%%%%%%%%%
\begin{figure}[ht!]
\begin{center}
\includegraphics[width=0.7\textwidth]{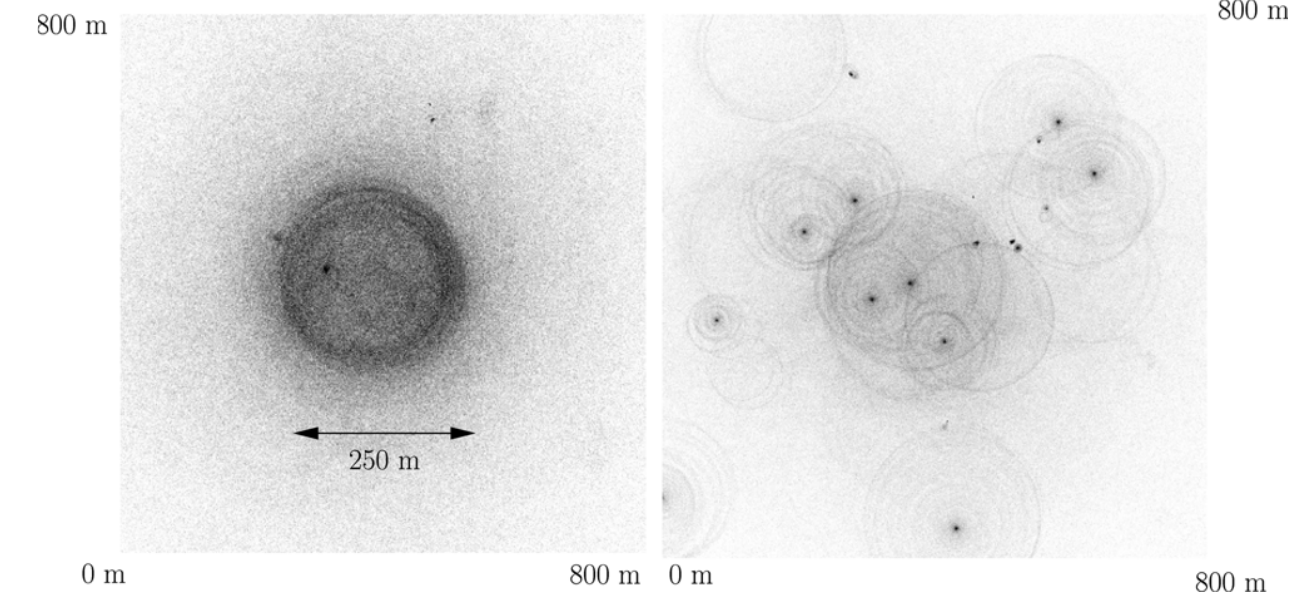}
\end{center}
\caption{Distribution of Cherenkov light on the ground from a 300 GeV $\gamma$-ray shower (left) and a 1 TeV proton shower (right) taken from Monte Carlo simulations \cite{aharonian2008}.}
\label{fig:cherenkov-images}
\end{figure}
%%%%%%%%%%%%%%%%%%%%%%%%%%%%%%%%%%%%%%%%%%%%%%%%%%%%%%%%%%%%%%%%%%
%

In 1978 Turver and Weekes with Monte Carlo calculations of $\gamma$-ray and proton-initiated air showers demonstrated for the first time that the Cherenkov imaging approach might provide a new technique for discrimination \cite{turver-weekes1978}. 
In fact, they observed that the lateral distribution of the light from gamma-ray and proton showers is quite different in appearance. Whereas the gamma-ray shower is a relatively compact light pool whose center is close to the shower axis, the proton shower is more scattered and has a wider range of fluctuations.
The scheme of the lateral development of air showers induced by protons (left side) and photons (right side) is shown in Fig. \ref{fig:cerenkov-scheme}. 
Images of EAS initiated by gamma rays have a compact elliptic shape, and the major axis of the ellipse indicates the shower axis projected onto the image plane. In contrast, the image of EAS produced by cosmic ray protons show a complex structure due to hadronic interactions that produce neutral pions (and electromagnetic subshowers initiated by the prompt decays of $\pi^0$ mesons to gamma rays) as well as penetrating muons from the decay of charged pions.

The different lateral structures induce different images on the pixel camera of a IACT.
The distribution of Cherenkov light on the ground from a 300 GeV $\gamma$-ray shower (left) and a 1 TeV proton shower (right) taken from Monte Carlo simulations is shown in Fig. \ref{fig:cherenkov-images}.

%
%%%%%%%%%%%%%%%%%%%%%%%%%%%%%%%%%%%%%%%%%%%%%%%%%%%%%%%%%%%%%%%%%%
\begin{figure}[ht!]
\begin{center}
\includegraphics[width=0.7\textwidth]{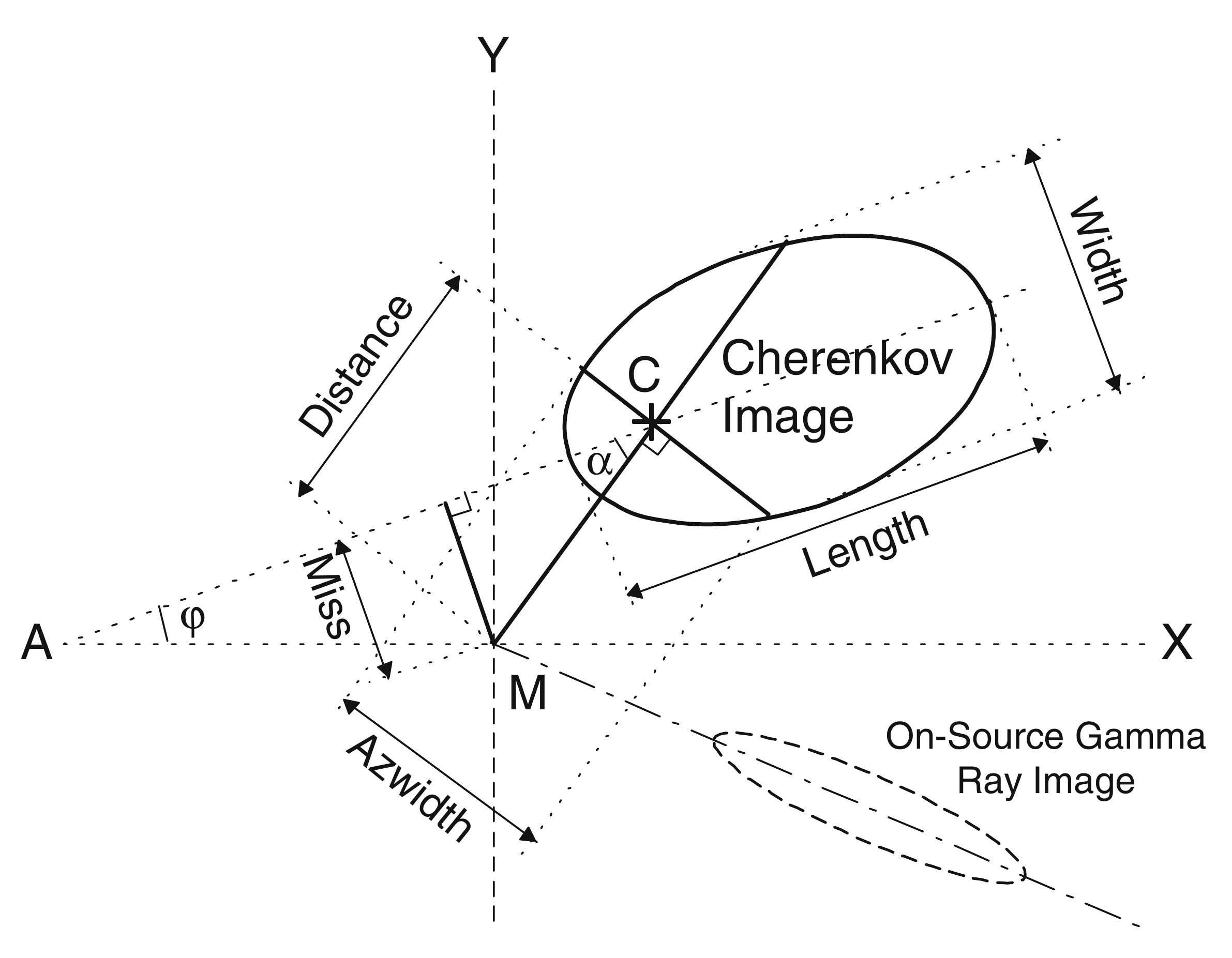}
\end{center}
\caption{Scheme of the elliptical image produced by an air shower on the pixel camera of a IACT. The different Hillas parameters, defined here, are used to perform background rejection.}
\label{fig:hillas-parameters}
\end{figure}
%%%%%%%%%%%%%%%%%%%%%%%%%%%%%%%%%%%%%%%%%%%%%%%%%%%%%%%%%%%%%%%%%%
%

In 1985 at the ICRC (La Jolla) Hillas suggested to use the so-called \emph{"Hillas image parameters"} to reduce the background: a key milestone in the history of Imaging Air Cherenkov Telescopes \cite{hillas1985}.
The idea was to quantify with some suitable parameters the observation that gamma showers are slimmer, more concentrated and orientated towards the source, as it can be seen from the Fig. \ref{fig:hillas-parameters}, where the Cherenkov footprint of a shower induced by photons on the focal plane of a Cherenkov telescope is sketched. 

The solid ellipse indicates the pixel image contour, C is the centroid of the image (location of highest brightness) and M the center of the field of view. 
The relevant parameters are the major and minor axis of the ellipse, labeled \emph{"Length"} and \emph{"Width"} in the plot, the angle $\alpha$ between the major axis and the line connecting the centroid C with the center of the field of view M, the \emph{"Distance"} between C and M, and the two quantities called \emph{"Miss"} and \emph{"Azwidth"}. 

Miss is the offset or the perpendicular distance between the extension of the major axis of the ellipse and M, and Azwidth is the azimuthal width of the image as indicated; it is the r.m.s. spread of light perpendicular to the line connecting C with M. 
Except for the clean regular elliptic shape this image is also representative for hadronic showers. 

The dashed ellipse at the lower right with the extension of the major axis intercepting the center M of the mirror, labeled \emph{"On-Source Gamma Ray Image"}, shows the typical narrow elliptic contour of a gamma ray shower when the mirror axis is pointing at the source and the impact parameter is non-zero.

Gamma rays are selected based on cuts on these parameters. Below, we briefly summarize the most important image characteristics used in IACT image analysis:
\begin{itemize}
\item \emph{Width}: the RMS angular size along the minor axis of the ellipse and related to the lateral development of the shower \emph{width} = $\sqrt{(\sigma_x^2 + \sigma_y^2 - z)/2}$.
\item \emph{Length}: the RMS angular size along the major axis of the ellipse and related to the longitudinal development of the shower \emph{length} = $\sqrt{(\sigma_x^2 + \sigma_y^2 + z)/2}$ to the centroid of the light distribution. This parameter gives a measure of the parallax angle to the shower max and grows with increasing impact parameter.
\item \emph{Alpha}: the angle made between the major axis of the ellipse and the line between the source position and centroid. Showers that originate from an object at the source position should have a value of alpha very close to zero. For an array of telescopes, rather than using alpha one typically uses the parameter $\theta^2$ that gives the square of the distance between the intersection point of the major axes of images in all pairs of telescopes and the source position. $\theta^2$ actually allows one, using stereoscopic arrays, to reconstruct the direction within 0.1 deg on an event-by-event base.
\end{itemize}
In the above definitions, we have used $d = \sigma_y^2 - \sigma_x^2$ and $z = \sqrt{d^2 + 4(\sigma_{xy}^2)^2}$. Differences in these parameters between gamma ray and proton showers can be used to distinguish gamma ray showers from the background events by cosmic ray particles with a single IACT.
For analysis of data from multiple telescopes, one typically extends this approach and derives weighted combinations of the width and length parameters \cite{aharonian2008}.

\begin{figure}[ht!]
\begin{center}
\includegraphics[width=0.7\textwidth]{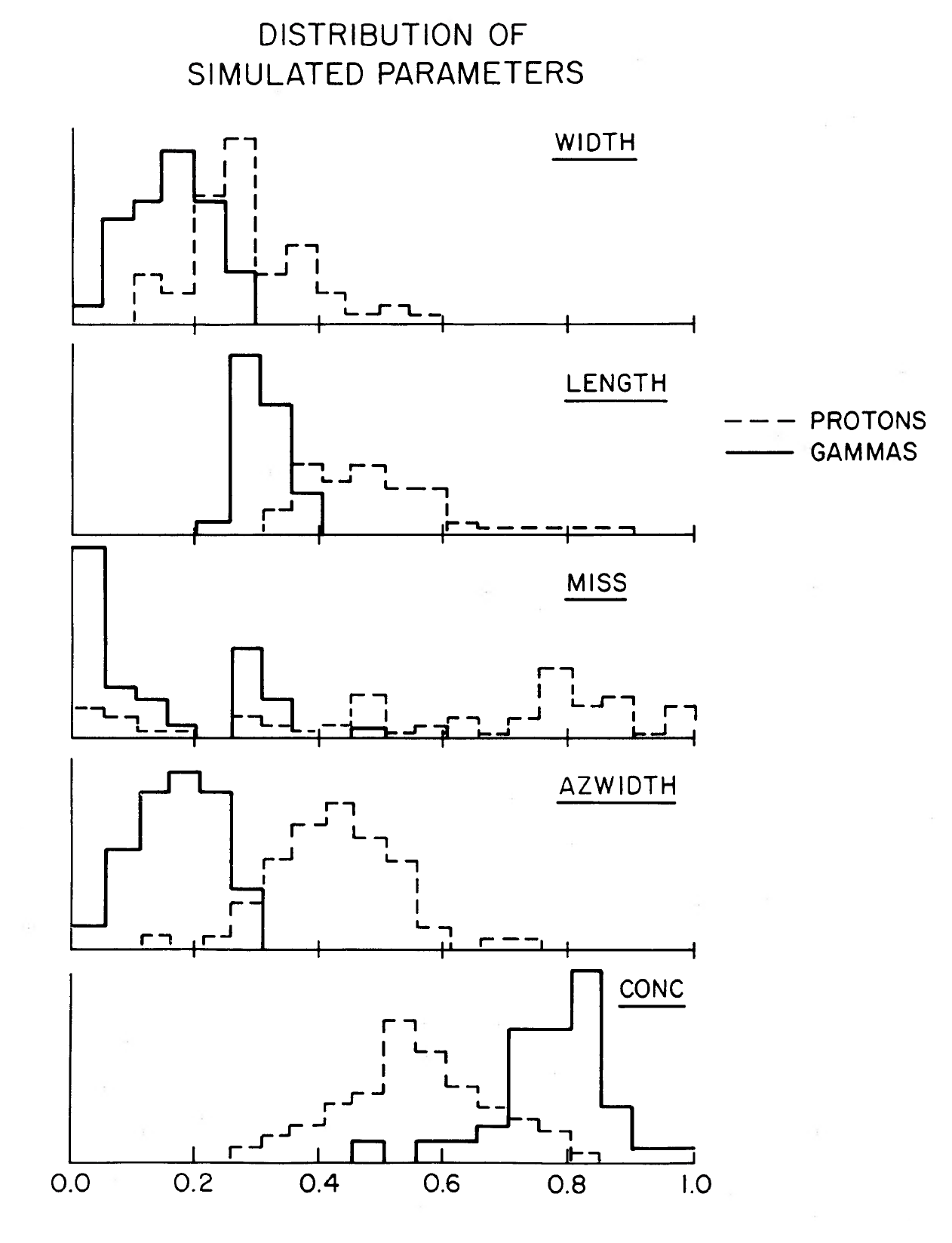}
\end{center}
\caption{Image paramenter distribution predicted by MonteCarlo simulations for an on-axis gamma-ray source at the zenith and an isotropic background of CRs \cite{whipple1989}.}
\label{fig:whipple-hparam}
\end{figure}

The real breakthrough in gamma-ray astronomy occurred in 1988 when the Whipple Collaboration applied the Hillas parameters to reject the background from data imaged by the 159 pixel camera of the 10 m Cherenkov telescope. They observed a gamma-ray signal in the direction of the Crab Nebula with a statistical significance of 9 standard deviations \cite{whipple1989}.
In Fig. \ref{fig:whipple-hparam} the distributions of simulated Hillas parameters for the Whipple telescope are shown \cite{whipple1989}. The difference between gamma-induced events and the isotropic background of charged CRs can be easily appreciated. With suitable cuts on these parameters the background can be enormously reduced.

The results obtained by the Whipple Collaboration marked the birthday of the ground-based VHE gamma astronomy.
%
%%%%%%%%%%%%%%%%%%%%%%%%%%%%%%%%%%%%%%%%%%%%%%%%%%%%%%%%%%%%%%%%%%
%%%%%%%%%%%%%%%%%%%%%%%%%%%%%%%%%%%%%%%%%%%%%%%%%%%%%%%%%%%%%%%%%%

New Cherenkov telescopes with improved background discrimination algorithms allow an effective rejection of hadronic showers by a factor of 100 making IACTs the ideal solution for ground-based gamma-ray astronomy.
Air shower array, in fact, cannot compete with IACTs in the identification of the background.
%
%%%%%%%%%%%%%%%%%%%%%%%%%%%%%%%%%%%%%%%%%%%%%%%%%%%%%%%%%%%%%%%%%%
\begin{figure}[ht!]
\begin{center}
\includegraphics[width=0.7\textwidth]{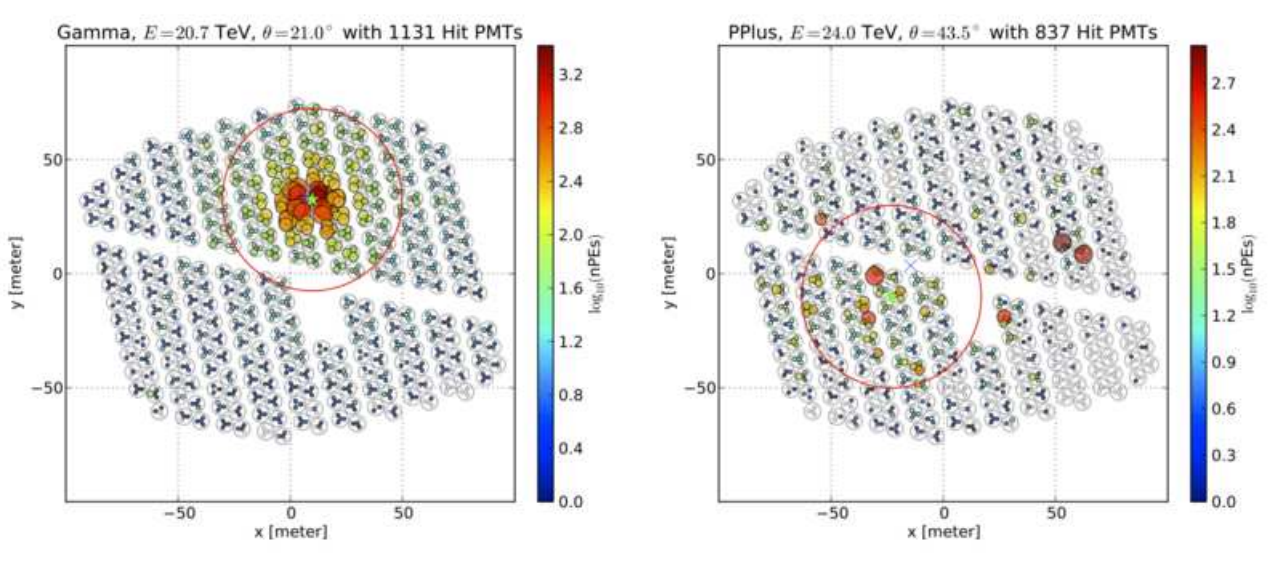}
\end{center}
\caption{Example for a simulated gamma shower (left panel) and proton shower (right panel) as observed by HAWC. The color code indicates the number of photoelectrons per PMT \cite{westerhoff2014}.}
\label{fig:hawc-gammaprot}
\end{figure}
%%%%%%%%%%%%%%%%%%%%%%%%%%%%%%%%%%%%%%%%%%%%%%%%%%%%%%%%%%%%%%%%%%

Arrays apply two different techniques to reduce the CR background:
\begin{enumerate}
\item Shower compactness.\\
Showers induced by charged CRs typically deposit large amounts of energy in distinct clumps far from the shower core ($>$40 m).
Experiments exploiting water Cherenkov technique (Milagro, HAWC, LHAASO) study the pattern of energy deposition in the detector and reject events induced by CRs by using topological cuts in the hit pattern (see Fig. \ref{fig:hawc-gammaprot}).
However, topology imply a minimum number of fired PMTs to have a pattern (HAWC is presently working with more than 70 PMTs), this means that the energy threshold is close to TeV.
The problem to discriminate the background below the TeV is still an open problem for shower arrays.
\item Muon content.\\
At high energy (above $\approx$10 TeV) events induced by photons can be distinguished from showers induced by charged nuclei by measuring their content of muons. In $\gamma$-ray showers muons are produced mainly by the pion photo-production process $\gamma$ + air $\to$ n$\pi^{\pm}$ + m$\pi^0$ + X, followed by the pion decays in muons and photons.
The probability of pion production, with respect to the probability to produce a e$^+$e$^-$ pair, is
\be
R = \frac{\sigma(\gamma - air \to \pi^+\pi^-)}{\sigma(\gamma - air \to e^+e^-)} \approx 3\cdot 10^{-3}.
\ee
This implies that the muon content in a gamma shower is only a few percent the muon content in a shower induced by charged CRs.
Detection of muons is not straightforward, however, the detection of the signal depends not only on the size and the sophistication of the detectors but also on the properties of the $\gamma$-ray showers. 
In fact, if these events have the same muon content as hadronic showers the detection of any $\gamma$-ray fluxes would be impossible.
In order to evaluate the rejection power it is crucial to study how frequently hadronic showers fluctuate in such a way to have a low muon content indistinguishable from $\gamma$-induced events.
The LHAASO experiment is installing a muon detector with unprecedented area ($\approx$40,000 m$^2$) allowing a discrimination of the background at a level of 10$^{-5}$ in the 100 TeV energy range.
\end{enumerate}

To better understand advantages and limits of the two techniques it is useful to examine the characteristics of a ground-based detector with particular attention to the parameters that define its sensitivity.

\section{Sensitivity to a $\gamma$-ray point source}

In gamma-ray astronomy, the ultimate characteristics of a detector is given by the sensitivity to a known point-source standard candle. 

The statistical significance of an observation is a key issue because it determines whether a given astronomical source has been detected or not, providing a probability for the excess being due to background fluctuations. It also plays an important role when the goal is to set upper limits for non-detected sources. In this case, the sensitivity of the method determines how constraining the upper limit is.

The main limitation of the ground-based $\gamma$-ray measurements is related to the difficulty into unambiguously identify and reject the charged cosmic ray background. This means that ground-based instruments detect a source as an excess of events from a certain direction over an overwhelming uniform background.
For this reason the capability to detect a photon signal over the background of charged cosmic rays can be simply expressed in units of standard deviations of the CR background, through the so-called \emph{"signal to noise ratio" S}:

The typical observation in $\gamma$-ray astronomy is following \cite{lima}.
A detector points in the direction of a suspected source, the so-called \emph{on-region}, for a certain time $t_{on}$ and counts N$_{on}$ particles. The counts in it are due to a possible source and the background.
The latter is determined by the count number N$_{off}$ in some \emph{off-region} for a time interval $t_{off}$ 
 It must be chosen in such a way that one can exclude a priori that it contains a source.
The quantity $\alpha$ is the ratio of the on-source time to the off-source time, $\alpha= t_{on}/t_{off}$.
Then we can estimate the number of background events included in the on-source counts  N$_{on}$: $\overline{N}_B = \alpha\> N_{off}$.
The observed signal, the probable number of photons contributed by the source, is
\be
N_S = N_{on} - \overline{N}_B = N_{on} - \alpha\cdot N_{off}
\ee
For a positive observation of an emission source, the excess counts $N_{on} - \overline{N}_B$ may have been caused only by a statistical fluctuation in the background rate.

There have been various procedures adopted by different experiments to estimate statistical reliability.
The so-called \emph{"Li\&Ma formula"} is the most frequently used method for estimating the significance of observations carried out in gamma-ray astronomy (eq. 17 in ref. \cite{lima}).

Some variant methods have been adopted to estimate the significances when experimenters reported their positive results of $\gamma$-ray sources.
For example, many experimenters use the standard deviation of the number of background particles N$_B$ as a measure of the statistical error of the observed signal N$_S$, and define the significance as
\be
S\frac{N_S}{\sqrt{\overline{N}_B}} = \frac{N_S}{\sqrt{\alpha\cdot N_{off}}}
\ee
As discussed in \cite{lima}, this formula overestimate the significance of the observation.
Nevertheless, to introduce the key parameters which determine the sensitivity of a ground-based $\gamma$-ray telescope we can use this approximate formula (with $\alpha$=1 and N$_{off}$=N$_{bkg}$)

\begin{equation}
\label{eq:sensit}
S=\frac{N_{\gamma}}{\sqrt{N_{bkg}}} = \frac{\int{J_{\gamma}(E)\cdot A_{eff}^{\gamma}(E)\cdot \epsilon_{\gamma}(E)\cdot f_{\gamma}(\Delta\Omega)\cdot T dE }}
{\sqrt{\int{J_{bkg}(E)\cdot A_{eff}^{bkg}(E)\cdot (1-\epsilon_{bkg}(E))\cdot \Delta\Omega\cdot T dE}}}
\end{equation}

where $J_{\gamma}$ and $J_{bkg}$ are the differential fluxes of photon and background, $A_{eff}^{\gamma}$ and $A_{eff}^{bkg}$ the effective areas, that determines the number of showers detected in a given observation time $T$, $\Delta\Omega=2\pi(1-cos\theta)$ the solid angle around the source and $ f_{\gamma}(\Delta\Omega)$ the fraction of $\gamma$-induced showers fitted in the solid angle.
The parameters $\epsilon_{\gamma}$ and $\epsilon_{bkg}$ are the efficiencies in identifying $\gamma$-induced and background-induced showers, respectively. As most of the parameters are function of the energy, the sensitivity depends on the energy spectra of the cosmic ray background and of the source.

The sensitivity S, formula (\ref{eq:sensit}), in 1 year can be expressed by

\begin{equation}
\label{eq:mdf}
S \propto \frac{\Phi_{\gamma}}{\sqrt{\Phi_{bkg}}}\cdot R\cdot \sqrt{A_{eff}^{\gamma}}\cdot \frac{1}{\sigma_{\theta}}\cdot {Q}
\end{equation}

where $\Phi_{\gamma}$ and $\Phi_{bkg}$ are the integral fluxes of photon and background, $\sigma_{\theta}$ is the angular resolution, $R=\sqrt{A_{eff}^{\gamma}/A_{eff}^{bkg}}$ the $\gamma$/hadron relative trigger efficiency and the so-called \emph{Q-factor} $Q=\frac{\epsilon_{\gamma}}{\sqrt{1-\epsilon_{bkg}}}$ represents the gain in sensitivity due to the background discrimination procedure.

For a point source the angular term to evaluate the background is given by the opening angle of the detector, i.e. the point spread function PSF ($\Delta\Omega = \Delta\Omega_{PSF} \sim \pi\, \theta^2_{PSF}$).
If we have an \emph{extended source} with a photon flux equal to that of the point source we must integrate over the extension of the source to have the same number of photons: $\Delta\Omega_{PSF}\to \Delta\Omega_{ext}$, and the background will increase.
Therefore, 
\be
S_{ext}\propto \bigg[ \frac{\Phi_{\gamma}}{\sqrt{\Phi_{bkg}}}\cdot R\cdot \sqrt{A_{eff}^{\gamma}}\cdot Q \cdot \frac{1}{\theta_{ext}} \cdot \frac{\theta_{ext}}{\theta_{PSF}}\bigg] \cdot \frac{\theta_{PSF}}{\theta_{ext}}
\ee
and
\be
S_{ext}\propto S_{point}\cdot \frac{\theta_{PSF}}{\theta_{ext}}
\ee
where $\theta_{ext}$ is the dimension of the extended source.
As it can be seen, detectors with a poor angular resolution, like shower arrays, are favoured in the extended source studies. 

Because for the integral fluxes we can write $\Phi_{\gamma}\sim E_{thr}^{-\gamma}$ and $\Phi_{bkg}\sim E_{thr}^{-\gamma_{bkg}}$ we obtain
\begin{equation}
\frac{\Phi_{\gamma}}{\sqrt{\Phi_{bkg}}}\sim E_{thr}^{-(\gamma - \gamma_{bkg}/2)} \sim E_{thr}^{-2/3}
\end{equation}
being $\gamma\sim$1.5 and $\gamma_{bkg} \sim$1.7.

Angular resolution, relative trigger probability, energy threshold and Q-factor are the main parameters, the drives, which determine the sensitivity of a ground-based $\gamma$-ray telescope.

\subsection{The energy threshold}

The energy threshold of EAS-arrays is not well defined. In fact, the trigger probability for a shower of a fixed energy increases slowly with energy mainly due to fluctuations in the first interaction height and is not a step function at the threshold energy $E_{thr}$.

The key to lower the energy threshold is to locate a detector at very high altitude. 
%
%%%%%%%%%%%%%%%%%%%%%%%%%%%%%%%%%%%%%%%%%%%%%%%%%%%%%%%%%%%
\begin{figure}[h!]
\begin{minipage}[t]{.47\linewidth}
  \centerline{\includegraphics[width=\textwidth]{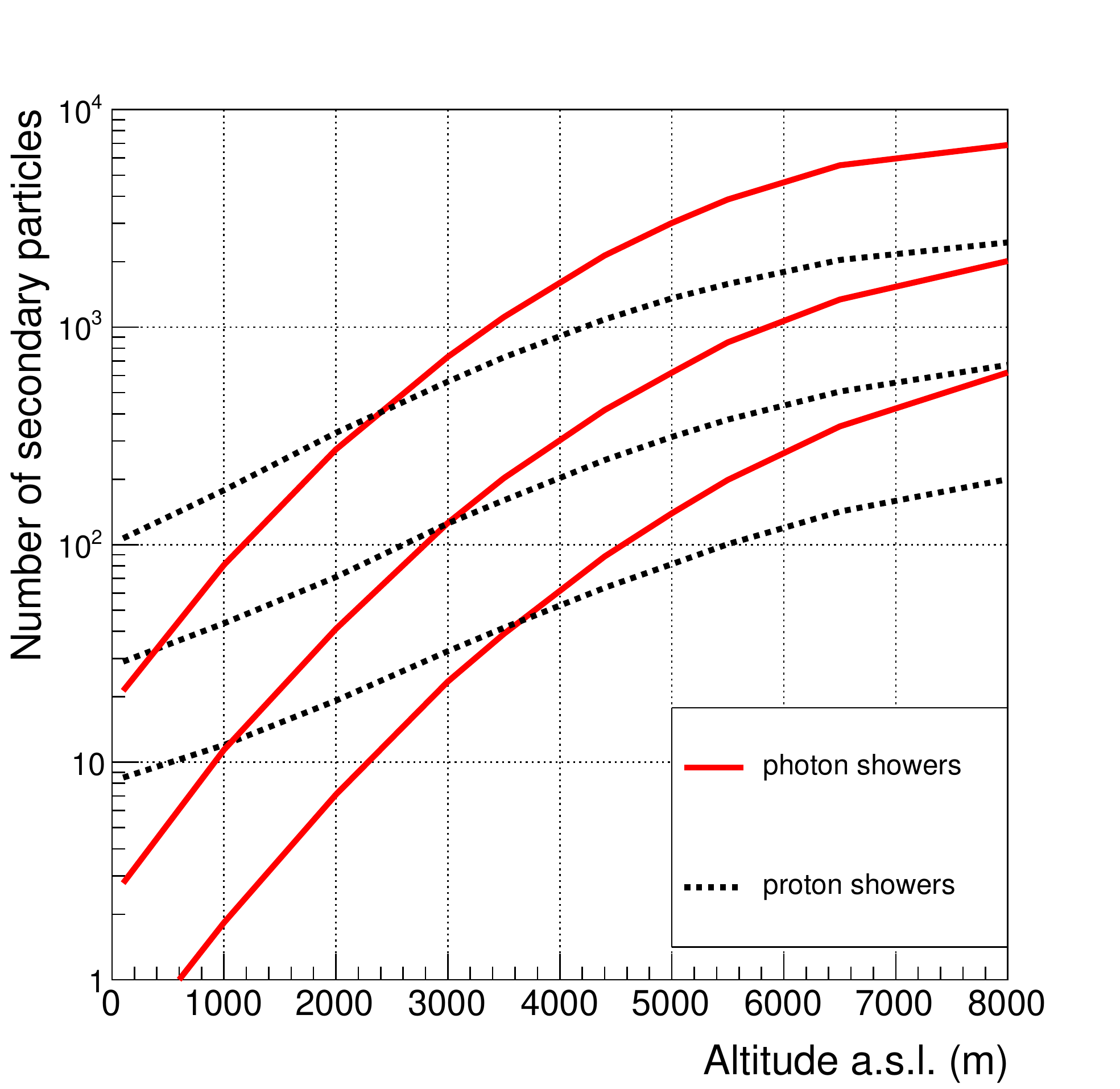} }
\end{minipage}\hfill
\begin{minipage}[t]{.47\linewidth}
  \centerline{\includegraphics[width=\textwidth]{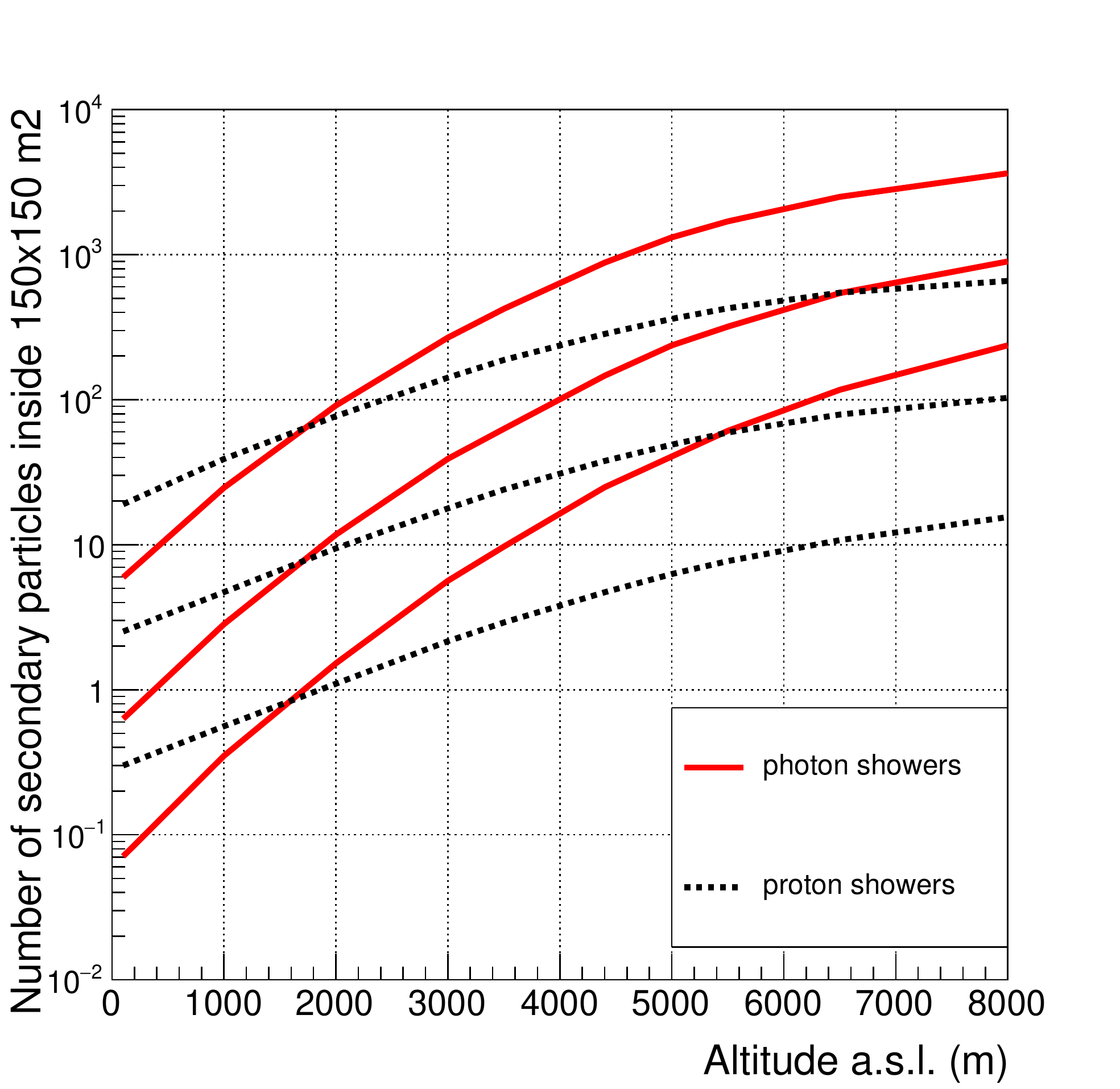} }
\end{minipage}\hfill
\caption[h]{Average number of particles (charged + photons) produced by showers induced by primary photons and protons of different energies at different observation levels. The left plot shows the total size, the right one refers to particles contained inside an area 150$\times$150 m$^2$ centered on the shower core. The plotted energies are 100, 300, 1000 GeV starting from the bottom \cite{disciascio-icrc2017}.} 
\label{fig:size}
\end{figure}
%%%%%%%%%%%%%%%%%%%%%%%%%%%%%%%%%%%%%%%%%%%%%%%%%%%%%%%%%%%
%
In the Fig. \ref{fig:size} the average sizes produced by showers induced by primary photons and protons of different energies at different observation levels are plotted. The left plot shows the total number of secondary particles (charged plus photons), the right one shows the number of particles contained inside an area 150$\times$150 m$^2$ centered on the shower core.
As can be seen, the number of particles in proton-induced events exceeds the number of particles in $\gamma$-induced ones at low altitudes. This implies that, in gamma-ray astronomy, the trigger probability is higher for the background than for the signal.

\begin{figure}[h!]
\begin{minipage}[t]{.47\linewidth}
  \centerline{\includegraphics[width=\textwidth]{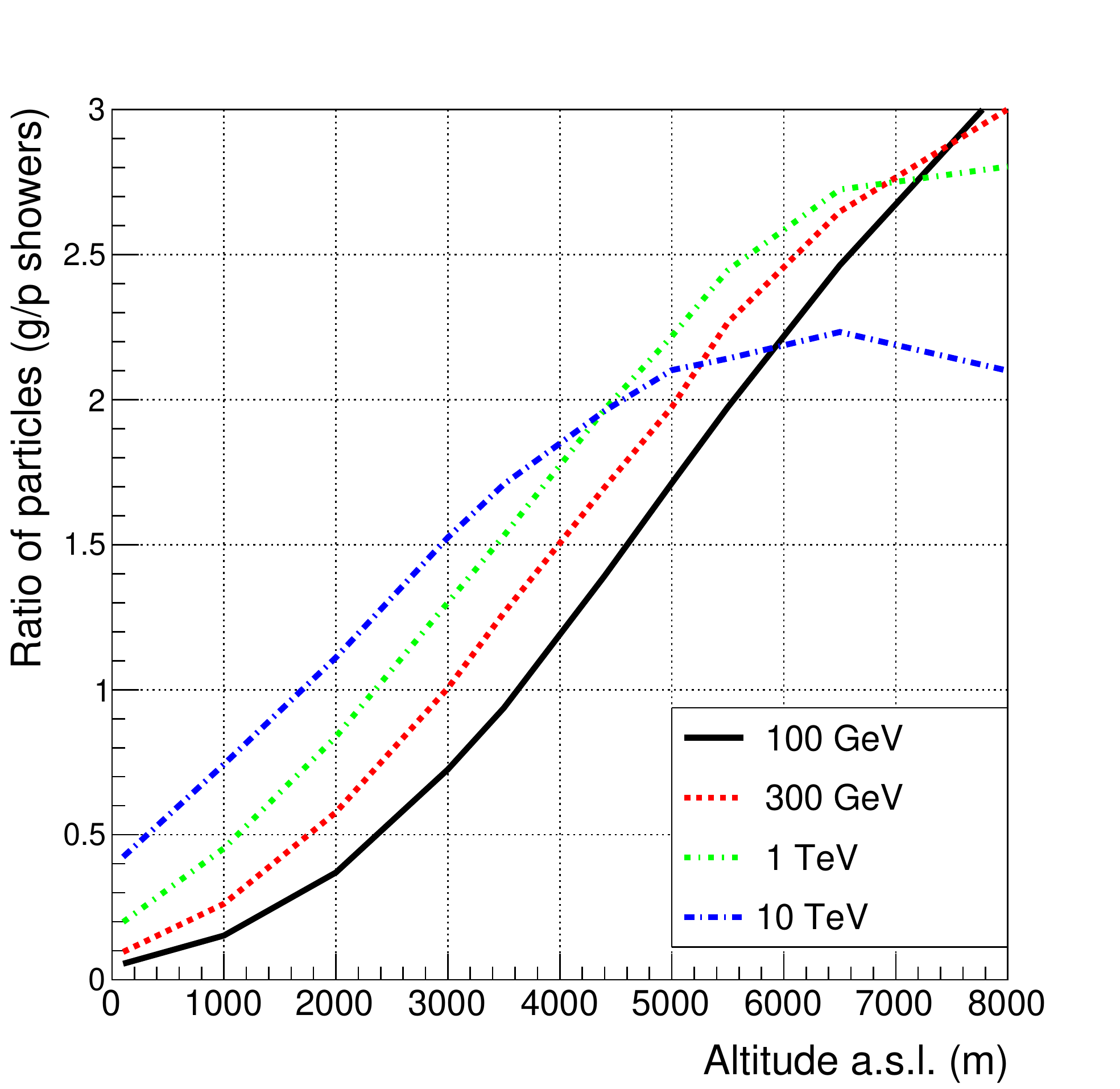} }
\end{minipage}\hfill
\begin{minipage}[t]{.47\linewidth}
  \centerline{\includegraphics[width=\textwidth]{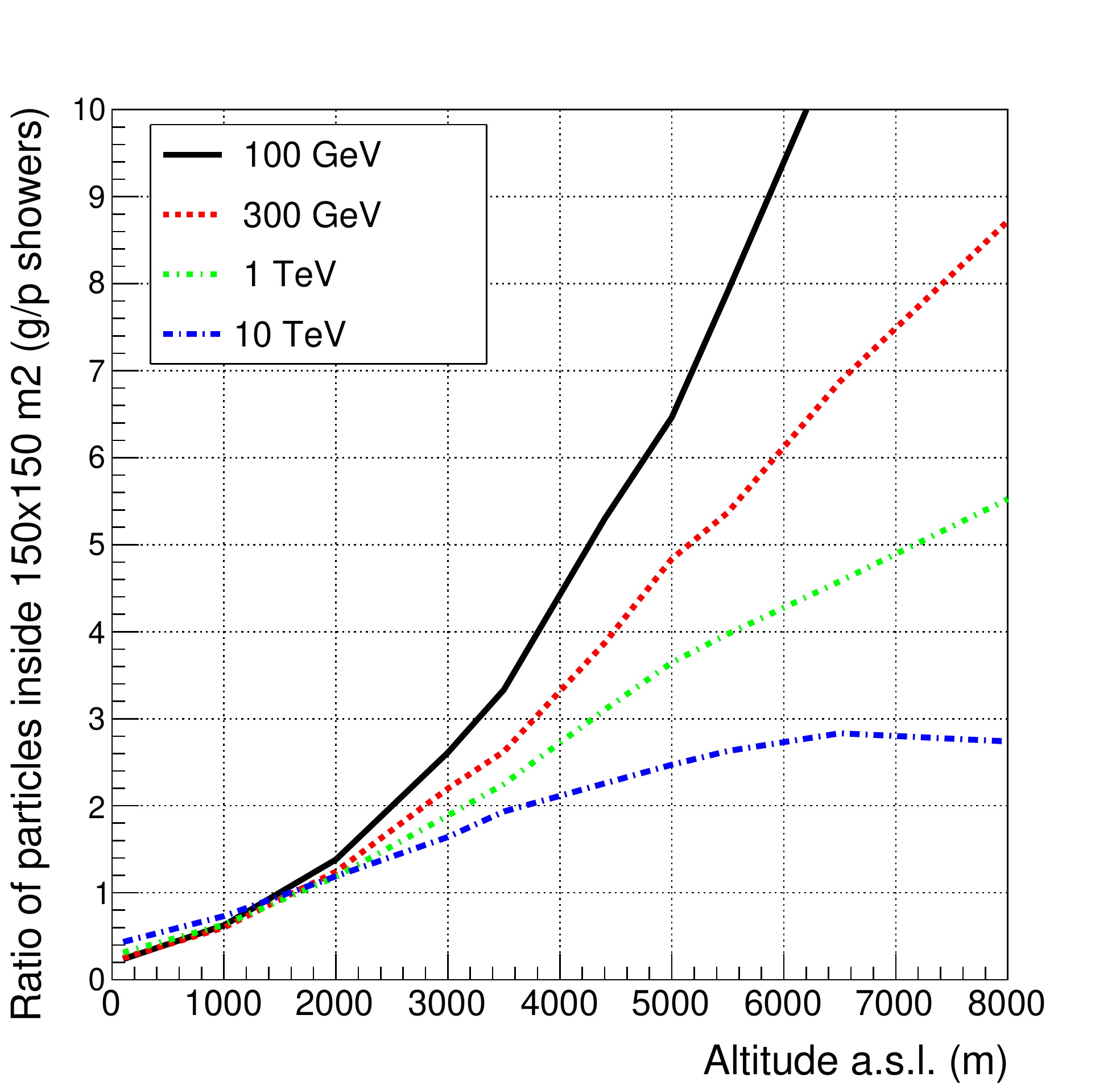} }
\end{minipage}\hfill
\caption[h]{Ratio of particle numbers (charged + photons) in photon- and proton-induced showers of different energies as a function of the observation level. 
The left plot shows the total size, the right one refers to particles contained inside an area 150$\times$150 m$^2$ centered on the shower core \cite{disciascio-icrc2017}.}
\label{fig:ratiosize}
\end{figure}

The small number of charged particles in sub-TeV showers within 150 m from the core imposes to locate experiments at extreme altitudes ($>$4500 m asl).
At 5500 m asl 100 GeV $\gamma$-induced showers contain about 8 times more particles than proton showers within 150 m from the core. This fact can be appreciated in the Fig. \ref{fig:ratiosize} where the ratio of particle numbers (charged + photons) in photon- and proton-induced showers of different energies as a function of the observation level are shown.

%
%%%%%%%%%%%%%%%%%%%%%%%%%%%%%%%%%%%%%%%%%%%%%%%%%%%%%%%%%%%
%%%%%%%%%%%%%%%%%%%%%%%%%%%%%%%%%%%%%%%%%%%%%%%%%%%%%%%%%%%
%

\subsection{Relative Trigger Efficiency R}

The effective area $A_{eff}$ is mainly a function of the number of charged particles at the observation level, the dimension and coverage of the detector and the trigger logic. Moving a given detector at different altitudes $A_{eff}$ is proportional to the number of charged particles.

The Fig. \ref{fig:ratiosize} shows the ratio R = N$^{\gamma}$/N$^p$ of all secondary particles (charged plus photons) in photon- and proton-induced showers as a function of the altitude. The left plot refers to the total size, the right one to particles contained inside an area 150$\times$150 m$^2$ centered on the shower core.
When R$<$1 the trigger probability is higher for protons than for photons. On the contrary, when R$>$1 the trigger probability, and thus the effective area, of the detector is larger for $\gamma$-showers than for protons, and an intrinsic $\gamma$/hadron-separation is available at higher altitudes. 
Being R proportional to the ratio of effective areas an altitude $>$4500 m asl is required to increase the sensitivity of a gamma-ray telescope in the hundreds GeV energy range.

Comparing the two plots of Fig. \ref{fig:ratiosize} we can see that, as expected, the $\gamma$-showers show a more compact particle distribution at the observation level. Therefore, at extreme altitudes the trigger efficiency of photon event at hundreds GeV is highly favoured if we consider only secondary particles within 150 m from the core. 
Showers of all energies have the same slope well after the shower maximum: $\approx$1.65x decrease per radiation length (r.l.). This implies that if a given detector is located one radiation length higher in atmosphere, the result will be a $\approx$1.65x decrease of the energy threshold.

But the energy threshold is also a function of the detection medium and of the coverage, the ratio between the detector and instrumented areas. Classical EAS arrays are constituted by a large number of detectors (typically plastic scintillators) spread over an area of order of 10$^4$ -- 10$^5$ m$^2$ with a coverage factor of about 10$^{-3}$. This poor coverage limits the energy threshold because small low-energy showers cannot be efficiently triggered by a sparse array. 
To exploit the potential of the coverage, a high granularity of the read-out must be coupled to image the shower front with high resolution. 

Another important factor to lower the energy threshold of a detector is the secondary photon component detection capability.
Gamma rays dominate the shower particles on ground: at 4300 m asl a 100 GeV photon-induced shower contains on average 7 times more secondary photons than electrons \cite{epas2}. 
In $\gamma$-showers the ratio N$_{\gamma}$/N$_{ch}$ decreases if the comparison is restricted to a small area around the shower core. For instance, we get N$_{\gamma}$/N$_{ch}\sim$3.5 at a distance r $<$ 50 m from the core for 100 GeV showers \cite{epas2}.

\begin{figure}[t!]
%\begin{minipage}[t]{.47\linewidth}
  \centerline{\includegraphics[width=0.7\textwidth]{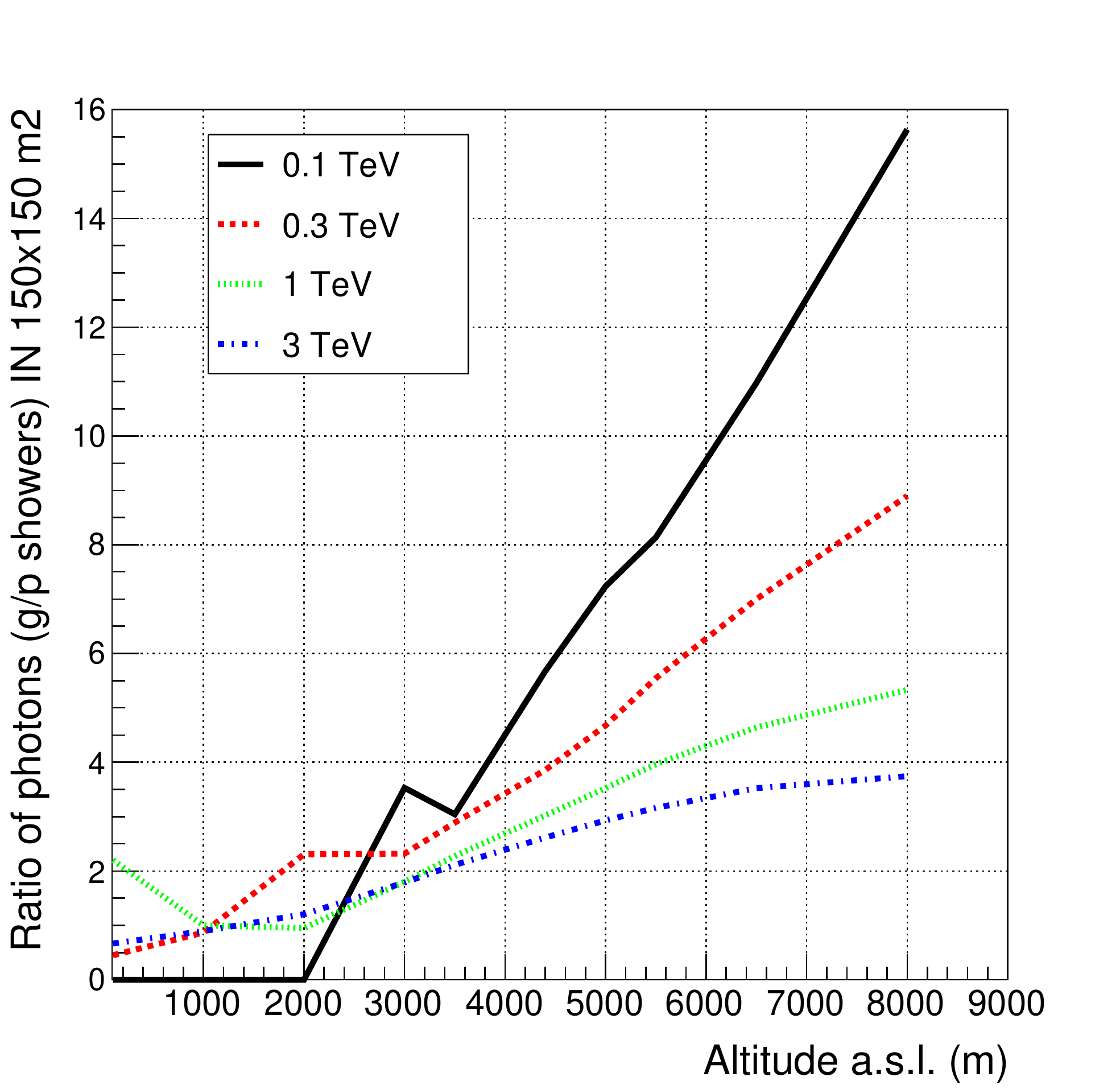} }
\caption[h]{Ratio of secondary photons in gamma- and proton-induced showers of different energies as a function of the observation level. The particles have been selected inside an area 150$\times$150 m$^2$ centered on the shower core \cite{disciascio-icrc2017}.} 
\label{fig:photons-pg}
%\end{minipage}\hfill
\end{figure}

In Fig. \ref{fig:photons-pg} the ratio of secondary photons within 150 m from the shower core for gamma- and proton-induced showers of different energies is plotted as a function of the altitude.
The number of secondary photons in low energy $\gamma$-showers exceeds by large factors the number of gammas in p-showers with increasing altitude. Detecting photons at extreme altitudes provides an intrinsic $\gamma$/hadron separation tool.
%
%%%%%%%%%%%%%%%%%%%%%%%%%%%%%%%%%%%%%%%%%%%%%%%%%%%%%%%%%%%
%%%%%%%%%%%%%%%%%%%%%%%%%%%%%%%%%%%%%%%%%%%%%%%%%%%%%%%%%%%
%

\subsection{The angular resolution}

In a search for cosmic point $\gamma$-ray sources with ground-based arrays the main problem is the rejection of the background due to charged cosmic rays, therefore a good angular resolution (i.e., the accuracy in estimating the arrival direction) is fundamental. 

The angular resolution $\sigma_{\theta}$ is related to the opening angle $\Delta \Omega$. If the point spread function of the angular resolution is gaussian $\sim e^{-\frac{\theta^{2}}{2\sigma_{\theta}^{2}}}$ the opening angle that maximize the $signal/bkg$ ratio is given by $\Delta \theta = 1.58 \sigma_{\theta}$ and it tallies with a fraction of $\epsilon=0.72$ of the events from the direction of the source in the solid angle $\Delta\Omega=2\pi (\-cos\Delta\theta)$. 
And so: $\frac{\epsilon(\Delta\Omega)}{\Delta\Omega}\simeq \frac{0.72}{1.6 \sigma_{\theta}}=\frac{0.45}{\sigma_{\theta}}$. In the following the opening angle $\Delta \Omega$ which maximize the sensitivity will be called $\psi_{70}$.

The usual method for reconstructing the shower direction is performing a $\chi^2$ fit to the recorded arrival times $t_i$ by minimization of 
\begin{equation}
\chi^2 = \sum_i w ( f - t_i )^2 
\end{equation}
where the sum includes all detectors with a time signal. Usually the function $f$ describes a plane, a cone with a fixed cone slope or a plane with curvature corrections as a function of core distance $r$ and multiplicity $m$. Fitted are a time offset and the two direction cosines. 
The weights $w$ are generally chosen to be an empirical function of the number $m$ of particles registered in a counter, a function of $r$ or a function of $r$ and $m$. This represents in general terms the usual fitting procedure of the "time of flight" technique.
Improvement to this scheme can be achieved by excluding from the analysis the time values belonging to the non-gaussian tails of the arrival time distributions by performing some successive $\chi^2$ minimizations for each shower \cite{fititer,argo-moon}.
In fact, the distribution of the arrival times shows non-Gaussian tails at later times, mainly due to multiple scattering of low energy electrons but also to incorrect counter calibrations and to random coincidences. These non-Gaussian tails are expected typically to be  20\% of all measured time values.

Placing a thin sheet of converter above the detector can improve the angular resolution due to, qualitatively:
(1) absorption of low energy electrons (and photons) which no longer contribute to the time signal;
(2) multiplication process of high-energy electrons (and photons) which produce an enhancement of the signal. The enhanced signal reduces the timing fluctuations: the contributions gained are concentrated near the ideal time profile because the high energy particles travel near the front of the shower while those lost tend to lag far behind.

\section{Imaging Atmospheric Cherenkov Telescopes}
\label{sect:iact}

The most successful instruments in the history of VHE $\gamma-$ray astronomy have been the Imaging Atmospheric Cherenkov Telescopes (IACTs). 
The imaging technique relies on the detection on the ground of the images of the Cherenkov light distribution from the electromagnetic cascades. From the measurement, it is possible to determine both the longitudinal and lateral development of the electromagnetic showers, and the arrival direction and energy of the primary $\gamma$-rays.
These detectors typically make use of a parabolic or spherical mirror to focus the Cherenkov photons onto a tightly packed array of photomultiplier tubes (PMTs) placed in the focal plane. 
A Cherenkov telescope must be operated in almost total darkness, in clear moonless nights. As a consequence, the duty cycle is limited to 10-15\%. 
Properties of the primary particles are inferred by the resulting image of the shower (see Fig. \ref{fig:cerenkov-scheme}).
%
%%%%%%%%%%%%%%%%%%%%%%%%%%%%%%%%%%%%%%%%%%%%%%%%%%%%%%%%%%%%%
\begin{figure}[ht!]
\begin{center}
\includegraphics[width=0.7\textwidth]{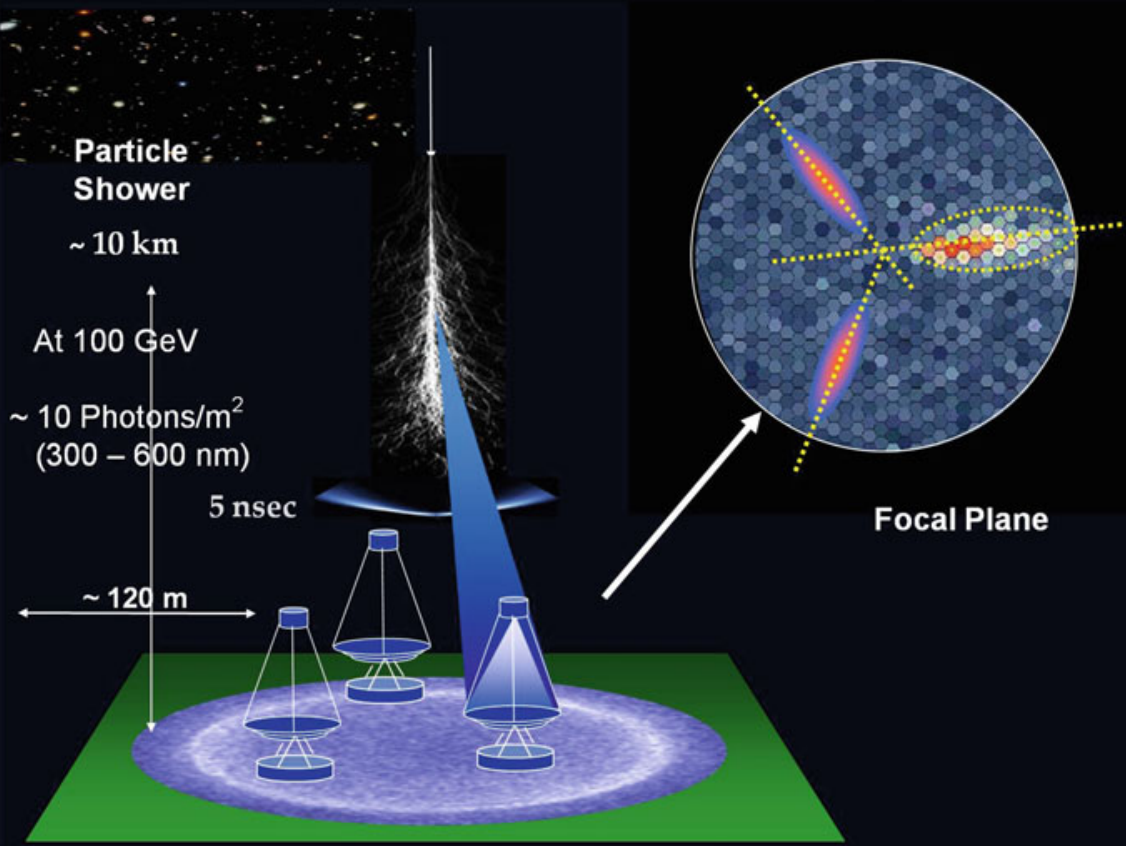}
\end{center}
\caption{Sketch of the imaging atmospheric Cherenkov telescopes technique. A shower initiated by a $\gamma$-ray of energy $\geqslant$100 GeV is observed through the Cherenkov radiation emitted by charged particles with $\sim$10 m diameter reflectors positioned about 100 m away from each other. The Cherenkov light arrives on the ground as a thin pancake, a few ns wide. The fraction arriving at the reflector is focused into a high-speed PMT camera in the focal plane. The images at the focal plane of different telescopes (upper right inset) allow for the determination of the shower's direction and energy \cite{spurio}.}
\label{fig:cerenkov-scheme-2}
\end{figure}
%%%%%%%%%%%%%%%%%%%%%%%%%%%%%%%%%%%%%%%%%%%%%%%%%%%%%%%%%%%%%
%

With a telescope consisting of an optical reflector of diameter D$\approx$ 10 m, as well as a camera with pixel size 0.1$^{\circ}$ - 0.2$^{\circ}$ and field of view $>$3$^{\circ}$, primary gamma-rays of energy $>$100 GeV can be collected from distances as large as 100 m. This provides huge detection areas, A$>$3$\times$10$^4$ m$^2$, which compensate the weak gamma-ray fluxes at these energies. The total number of photons in the registered Cherenkov light image is a measure of energy, the orientation of the image correlates with the arrival direction of the gamma-ray, and the shape of the image contains information about the origin of the primary particle (a proton or photon) (see for a review Ref. \cite{spurio}). The basic principles of operation of the IACT technique are illustrated in Fig. \ref{fig:cerenkov-scheme-2}.

%
%%%%%%%%%%%%%%%%%%%%%%%%%%%%%%%%%%%%%%%%%%%%%%%%%%%%%%%%%%%
\begin{figure}[ht!]
  \centerline{
  \includegraphics[width=0.8\textwidth]{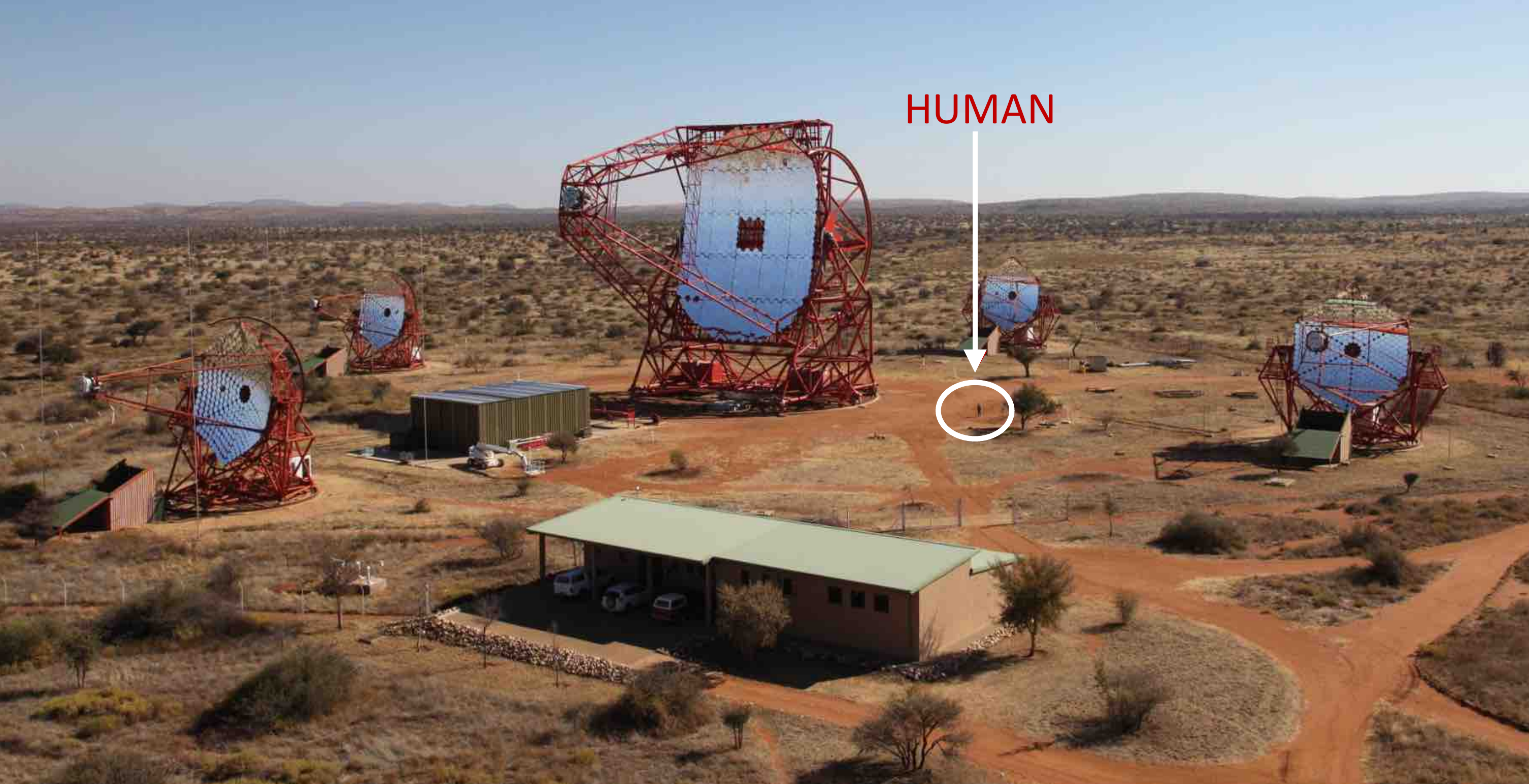}   }
\caption[h]{The HESS telescopes in Namibia.} 
\label{fig:hess}
\end{figure}
\begin{figure}[ht!]
  \centerline{\includegraphics[width=0.8\textwidth]{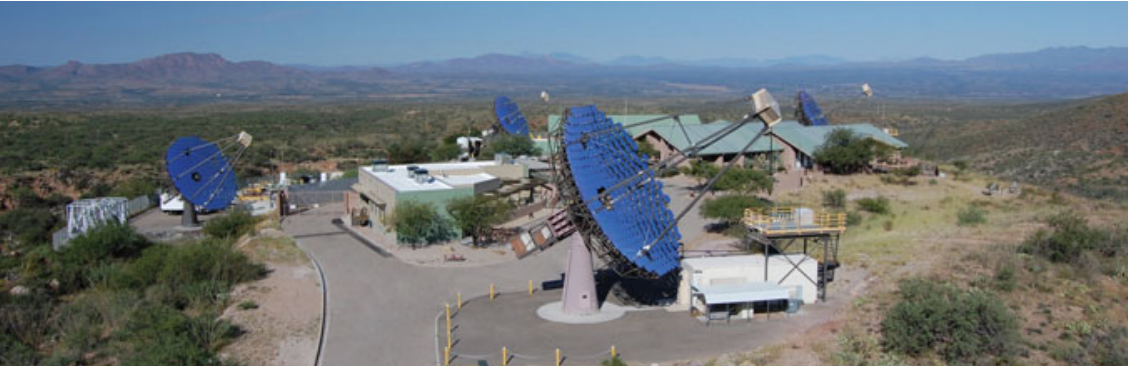} }
\caption[h]{The four IACT array VERITAS at Mt. Hopkins, Arizona.}
\label{fig:veritas}
\end{figure}
\begin{figure}[ht!]
  \centerline{\includegraphics[width=0.8\textwidth]{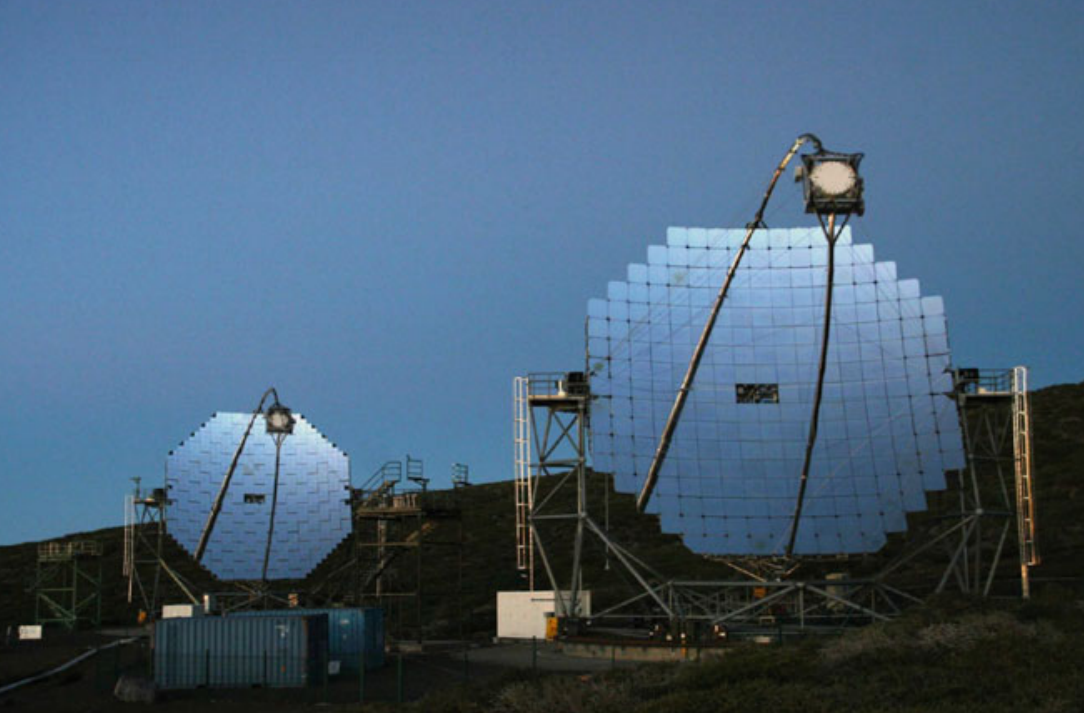} }
\caption[h]{The two large telescopes of the MAGIC observatory.}
\label{fig:magic}

\end{figure}
%%%%%%%%%%%%%%%%%%%%%%%%%%%%%%%%%%%%%%%%%%%%%%%%%%%%%%%%%%%
%
The stereoscopic observations of air showers with two or more 10 m diameter telescopes located at distances of about 100 m from each other provide a quite low energy threshold around 100 GeV, effective (by a factor of 100) rejection of hadronic showers, and good angular ($\approx$0.1$^{\circ}$) and energy ($\approx$15\%) resolutions. At energies around 1 TeV, the performance achieves the best sensitivity, a minimum detectable energy flux of 10$^{-13}$ erg/cm$^2$/s. This is a quite impressive achievement even in the standards of advanced branches of astrophysics \cite{aharonian-casanova}.

As discussed, their excellent results are actually mainly due to their ability to perform an efficient $\gamma$/hadron separation using the shape of the Cherenkov image \cite{hillas1985,whipple1989}: $\gamma-$rays give a smooth, elliptical image that points toward the centre of the field of view, while CRs images are broader and more irregular. Actually the background rejection makes indirectly use of the direction reconstruction: requiring that the axis of the elliptical image intersects the centre of the FoV is equivalent to the selection of showers whose direction is parallel to the mirror axis (i.e. primary particles coming from the pointed direction). In fact the background rejection is not so good in the case of diffuse sources. 

There are currently three major IACT systems in operation, two in the Northern hemisphere, and one in the Southern \cite{spurio}.
\begin{itemize}
\item The High Energy Stereoscopic System (HESS) Observatory.\\
HESS is located in Namibia (-23$^{\circ}$ N, -16$^{\circ}$ W, altitude 1800 m), in the Southern Hemisphere (see Fig. \ref{fig:hess}). 
It is the IACT with the largest field-of-view and the only one in the Southern hemisphere able to observe the Inner Galaxy and the Galactic Center.
 The initial four HESS telescopes (Phase I, completed in 2004) are arranged in the form of a square having a side length of 120 m, to provide multiple stereoscopic views of air showers. Each telescope of Phase I has a diameter of 13 m, with a total mirror area of 108 m$^2$ per telescope, 960 photon detector elements ("pixels") to resolve image details and a field-of-view of about 5$^{\circ}$. 
In Phase II of the project, a single huge dish with about 600 m$^2$ mirror area was added at the center of the array, increasing the energy coverage, sensitivity and angular resolution of the instrument. The camera of the new telescope contains 2048 pixels. This telescope started taking data in 2012.

\item The Very Energetic Radiation Imaging Telescope Array System (VERITAS).\\
VERITAS is in operation at the Fred Lawrence Whipple Observatory in southern Arizona, USA (32$^{\circ}$ N, 111$^{\circ}$ W, altitude 1275 m) (see, Fig. \ref{fig:veritas}). It is an array of four 12 m optical reflectors with similar characteristics as HESS-Phase I. Each reflector uses 499-pixel cameras, with a field-of-view of 3.5$^{\circ}$. The covered energy range is between 50 GeV and 50 TeV. The four-telescope array was completed in January 2007.

\item The Major Atmospheric Gamma-ray Imaging Cherenkov (MAGIC).
MAGIC (28$^{\circ}$ N, 17$^{\circ}$ W, altitude 2225 m) originally consisted of a single, very large reflector (236 m$^2$) installed on the Canary island of La Palma, with a 3.5$^{\circ}$ high-resolution camera composed of 576 ultra-sensitive PMTs (see Fig. \ref{fig:magic}). The first telescope has been fully operational since 2004. In 2009, a second telescope of essentially the same characteristics was added; MAGIC-II was installed at a distance of 85 m from MAGIC-I.
MAGIC is characterized by the largest collection surface of any existing $\gamma$-ray telescope worldwide, constituted by nearly 1000 individual mirrors, together resulting in a parabolic dish with a 17 m diameter.
The large surface and the optimal light collection of the mirrors allow for the detection of $\gamma$-rays with an energy threshold of about 20 GeV.
A very fast (average time 40 s) repositioning of the telescope axis is one of the major characteristics of MAGIC with respect to other IACTs. Repositioning in a matter of seconds is important when short-lived phenomena are triggered by other active devices, e.g., by satellite-based wide-angle detectors in the X-ray band, in particular, for GRBs. 

In fact, MAGIC for the first time observed from ground, with high statistical significance, the afterglow of a GRB, the GRB 190114C triggered by the Swift-BAT alert. The MAGIC real-time analysis shows a significance $>$20 standard deviations in the first 20 min of observations (starting at T0+50 s) for energies $>$300 GeV. The relatively high detection threshold is due to the large zenith angle of observations ($>$60 degrees) and the presence of partial Moon \cite{magic-grb}.  
\end{itemize}

%
%%%%%%%%%%%%%%%%%%%%%%%%%%%%%%%%%%%%%%%%%%%%%%%%%%%%%%%%%%%
\begin{figure}[ht!]
  \centerline{\includegraphics[width=0.7\textwidth]{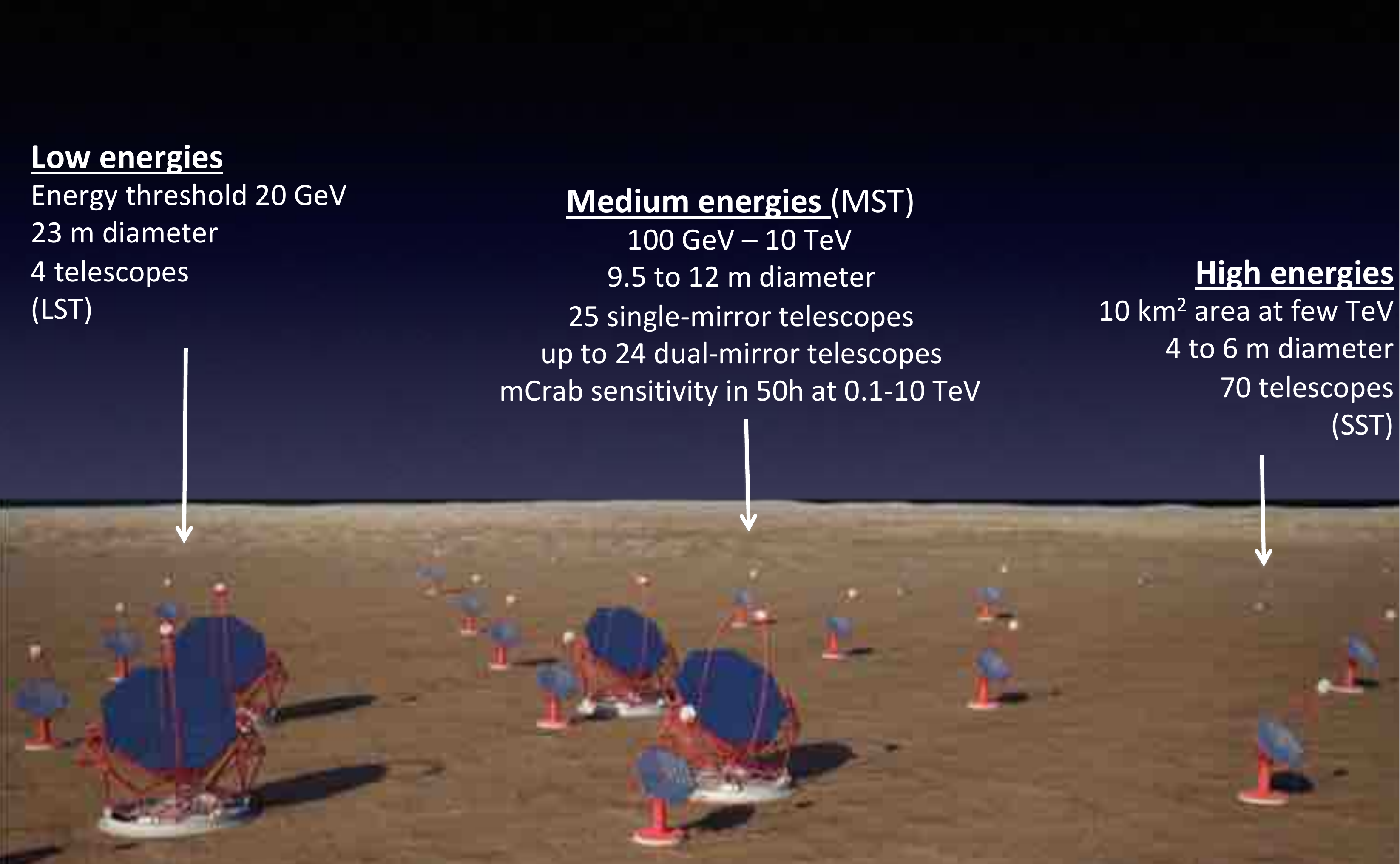} }
\caption[h]{Layout of the future CTA-South array.}
\label{fig:cta}
\end{figure}
%%%%%%%%%%%%%%%%%%%%%%%%%%%%%%%%%%%%%%%%%%%%%%%%%%%%%%%%%%%
%
Regarding the previous generation instruments, one should mention two detectors, the 10 m diameter single dish of the Whipple Observatory (south Arizona) and the HEGRA array of five relatively modest (4 m diameter) Cherenkov telescope (La Palma, Canary Islands). These instruments can be considered as prototypes of the current IACT arrays. They played a crucial role in the development of ground-based gamma-ray astronomy. While the Whipple collaboration, as discussed, pioneered the implementation and successful realization of the \emph{imaging} atmospheric Cherenkov technique \cite{whipple1989}, the HEGRA collaboration convincingly demonstrated the power of the \emph{stereoscopic} approach \cite{hegra-iact,hegra-iact2}.

The Cherenkov Telescope Array (CTA) will represent the future of observational $\gamma$-ray astronomy from ground (Fig. \ref{fig:cta}). 
The research objectives of the next generation of IACT arrays, and especially of CTA, are devoted to (1) a significant improvement (by an order of magnitude) of the flux sensitivities in the standard 0.1-10 TeV energy interval, and (2) an expansion of the energy domain of IACT arrays in both directions, down to 10 GeV and well beyond 50 TeV. This ambitious research goal will be realized by increasing the number of telescopes with different sizes, from 4 very large 20 m diameter class telescopes to a large number of modest area (10-30 m$^2$) reflectors \cite{cta-science} . 

The IACT arrays are designed to observe point-like or moderately extended (with angular size 1$^{\circ}$ or less) objects with known celestial coordinates. The potential of IACT arrays is limited for the search for very extended structures (like diffuse emission, Fermi Bubbles and giant Radio Lobes of the radiogalaxy Centaurus A), as well as for solitary or transient $\gamma$-ray phenomena. In this regard, the detection technique based on direct sampling of secondary particles that constitute the extensive air showers, is a complementary approach to the IACT technique.

\section{Extensive Air Shower Arrays}
\label{sect:array}

Instruments that detect the secondary charged particles reaching the ground level are known as EAS arrays. As discussed in the lecture \cite{disciascio-crs}, typically they consist of a number of charged particle detectors, usually 1 m$^2$ scintillation counters (or water Cherenkov tanks), spread over an area of 10$^4$ - 10$^5$ m$^2$ with a spacing of 10 - 20 m. The total sensitive area is therefore less than 1\% of the total enclosed area. The shower is sampled at a single depth (the observational level) and with an additional sampling of the shower front at fixed points. This results in a high degree of uncertainty in the reconstruction due to fluctuations\footnote{For a comparison, Cherenkov detectors record information related to the whole longitudinal and lateral development of the shower summarized by the elliptical image on the pixel camera.}.

The sparse sampling sets the energy threshold and determines a poor energy resolution ($\sim$100\%). The direction of the incoming primary particle is reconstructed with the fast timing method making use of the relative times at which the individual detection units are fired by the shower front. Also the angular resolution is limited by the shower fluctuations ($\sim 1^{\circ}$). In few arrays the measurement of the muon content may allow for background rejection above $\approx$10 TeV. However, the ability to discriminate $\gamma$ and hadron induced showers is quite limited.

In addition dependence of threshold and reconstruction capabilities on the zenith angle is higher than for Cherenkov detectors: since the active area is horizontal, its projection onto a plane perpendicular to the shower axis is smaller than the geometrical area\footnote{Cherenkov telescopes point the source therefore their area is always orthogonal to the shower axis. The degradation of the detector performance with zenith is simply due to the increasing thickness of atmosphere.}. On the other hand, EAS arrays have a large field of view ($\sim$2 sr) and a 100\%
duty cycle. These characteristics give them the capability to serve as all-sky monitors.

Two different experimental techniques have been applied in the last two decades in ground-based survey instruments: 
\begin{enumerate}
\item[(1)] Water Cherenkov (Milagro).
\item[(2)] Resistive Plate Chambers (ARGO-YBJ). 
\end{enumerate}

The Milagro detector consisted of a large central water reservoir (60$\times$80 m$^2$), which operated between 2000 and 2008 in New Mexico (36$^{\circ}$ N,107$^{\circ}$ W), at an altitude of 2630 m. The reservoir was covered with a light-tight barrier, and instrumented with 2 layers of 8'' PMTs to improve the detection of muons. In 2004, an array of 175 small tanks was added, irregularly spread over an area of 200$\times$200 m$^2$ around the central reservoir. With this array Milagro developed analysis techniques for CR background discrimination.

The ARGO-YBJ experiment consisted in a full coverage central carpet ($\sim$75$\times$75 m$^2$ and coverage $\sim$92\%) enclosed by a partially instrumented guard-ring to improve reconstruction of events with the core outside the carpet \cite{disciascio-rev}. The active elements of the detector were constituted by Resistive Plate Chambers (RPCs). ARGO-YBJ has been in stable data taking between 2007 and 2013 in Tibet at an altitude of 4300 m.
The benefits in the use of RPCs in ARGO-YBJ were \cite{disciascio-rev}: (1) high efficiency detection of low energy showers by means of the high density sampling of the central carpet (median energy in the first multiplicity bin $\sim$300 GeV); (2) unprecedented wide energy range investigated by means of the digital/charge read-outs ($\sim$300 GeV $\to$ 10 PeV); (3) good angular resolution ($\sigma_{\theta}\approx 1.66^{\circ}$ at the threshold, without any lead layer on top of the RPCs) and unprecedented details in the core region by means of the high granularity of the different read-outs.
RPCs allowed to study also charged CR physics (energy spectrum, elemental composition and anisotropy) up to about 10 PeV.
By contrast, the capability of water Cherenkov facilities in extending the energy range to PeV and in selecting primary masses must be still demonstrated.

In both experiments (Milagro and ARGO-YBJ) the limited capability to discriminate the background was mainly due to the small dimensions of the central detectors (pond and carpet). In fact, in the new experiments, like HAWC and LHAASO, the discrimination of the CR background is made studying shower characteristics far from the shower core (at distances R$>$ 40 m from the core, the dimension of the Milagro and ARGO-YBJ detectors). Milagro developed analysis techniques for CR background discrimination only when the array of small tanks was added and the instrumented area enlarged.

The ARGO-YBJ experiment, combining the full coverage approach at high altitude with a high granularity of the read-out (about 15,000 strips 7$\times$62 cm$^2$ wide), sampled 100 GeV $\gamma$-induced showers with an efficiency of about 70\%. 
The median energy of the first multiplicity bin (20-40 fired pads) for photons with a Crab-like energy spectrum was 340 GeV \cite{argo-crab}.
The granularity of the read-out at cm level allowed to sample events with only 20 fired pads, out of 15,000, with a background-free topological-based trigger logic.
A water Cherenkov based facility will hardly be able to lower the energy threshold below $\sim$500 GeV. 
%
%%%%%%%%%%%%%%%%%%%%%%%%%%%%%%%%%%%%%%%%%%%%%%%%%%%%%%%%%%
\begin{figure}[ht!]
\centerline{\includegraphics[width=0.8\textwidth]{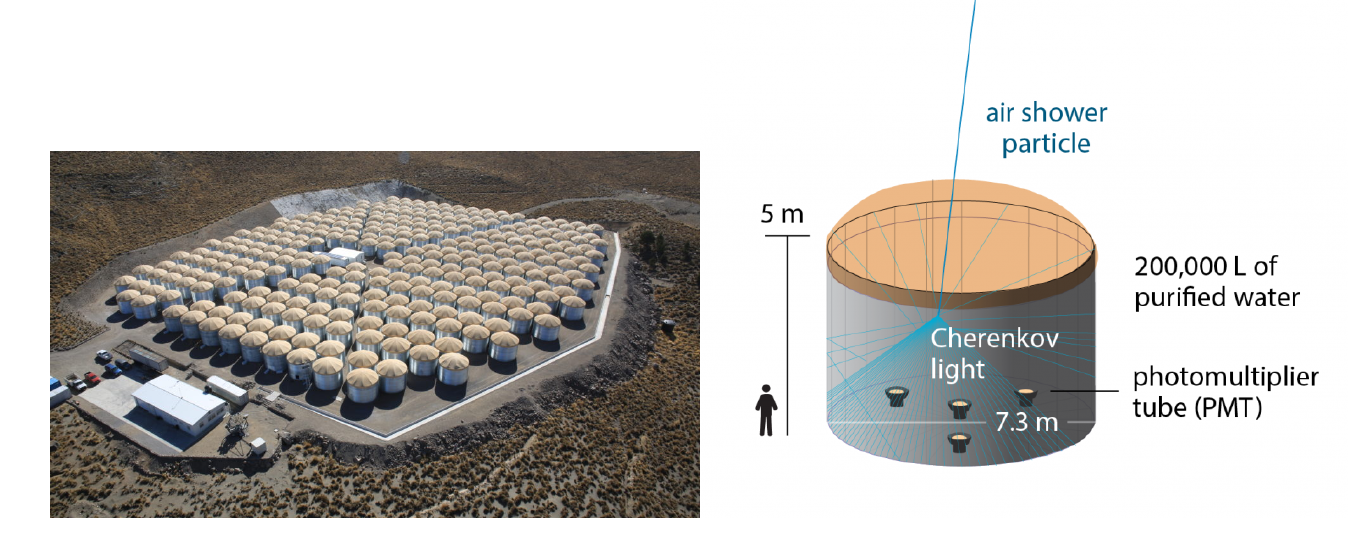} }
\caption{Layout of the HAWC experiment. The sketch of a water tank is shown on the right side.} 
\label{fig:hawc-layout}       % Give a unique label
\end{figure}
%%%%%%%%%%%%%%%%%%%%%%%%%%%%%%%%%%%%%%%%%%%%%%%%%%%%%%%%%%
%
The Milagro results, as well as the potential for continuous monitoring of a large fraction of the sky, have motivated the construction of larger EAS detectors like the High-Altitude Water Cherenkov Observatory (HAWC). 
HAWC is located on the Sierra Negra volcano in central Mexico at an elevation of 4100 m a.s.l.  (see Fig. \ref{fig:hawc-layout}).
It consists of an array of 300 water Cherenkov detectors made from 5 m high, 7.32 m diameter, water storage tanks  covering an instrumented area of about 22,000 m$^2$ (the actual tank coverage is 12,550 m$^2$ with a coverage factor less than 60\%).
Four upward-facing photomultiplier tubes (PMTs) are mounted at the bottom of each tank: a 10'' PMT positioned at the center, and three 8'' PMTs positioned halfway between the tank center and rim. 
The central PMT has roughly twice the sensitivity of the outer PMTs, due to its superior quantum efficiency and larger size. The WCDs are filled to a depth of 4.5 m, with 4.0 m (more than 10 radiation lengths) of water above the PMTs. This large depth guarantees that the electromagnetic particles in the air shower are fully absorbed by the HAWC detector, well above the PMT level, so that the detector itself acts as an electromagnetic calorimeter providing an accurate measurement of e.m. energy deposition \cite{hawc12,hawc1,hawc2}.

After a few years of operations a number of interesting results has been obtained.
We mention just the publication of the 2nd catalog of gamma-ray sources \cite{hawc-2ndcat} that demonstrates the enormous potential of wide FoV detectors in discovering new $\gamma$-ray sources.
Realized with 507 days of data, with an instantaneous FoV $>$1.5 sr and $>$90\% duty cycle, represents the most sensitive TeV survey to date for such a large fraction of the sky. 
The median energy of the lowest multiplicity bin is about 700 GeV. 
A total of 39 sources were detected, with an expected contamination of 0.5 due to background fluctuations. Out of these sources, 19 are new sources that are not associated with previously known TeV sources. Ten are reported in TeVCat as PWN or SNR: 2 as blazars and the remaining eight as unidentified.

Two new wide FoV detectors, LHAASO and TAIGA-HiSCORE, are under installation and will be briefly described in the following.
%
%%%%%%%%%%%%%%%%%%%%%%%%%%%%%%%%%%%%%%%%%%%%%%%%%%%%%%%%%%
\begin{figure}[ht!]
\centerline{\includegraphics[width=0.8\textwidth]{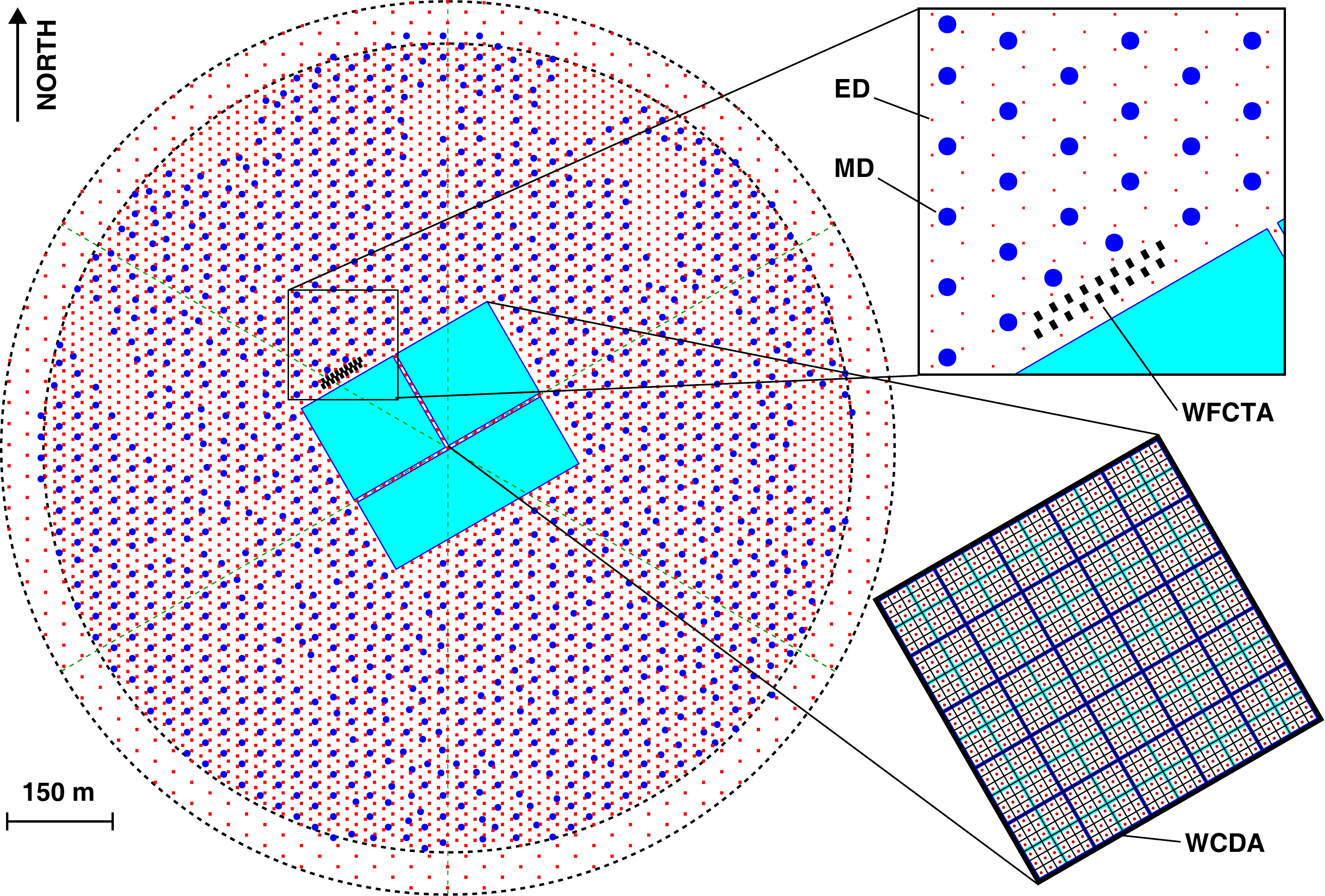} }
\caption{Layout of the LHAASO experiment. The insets show the details of one pond of the WCDA and of the KM2A array constituted by two overimposed arrays of electromagnetic particle detectors (ED) and of muon detectors (MD). The telescopes of the WFCTA, located at the edge of a pond, are also shown.} 
\label{fig:lhaaso-layout}       % Give a unique label
\end{figure}
%%%%%%%%%%%%%%%%%%%%%%%%%%%%%%%%%%%%%%%%%%%%%%%%%%%%%%%%%%
%

\subsection{The LHAASO experiment}

A new project, developed starting from the experience of ARGO-YBJ,  is LHAASO \cite{lhaaso1}. 
The experiment is strategically built to study with unprecedented sensitivity the energy spectrum, the elemental composition and the anisotropy of CRs in the energy range between 10$^{12}$ and 10$^{17}$ eV, as well as to act simultaneously as a wide aperture ($\sim$2 sr), continuosly-operated gamma-ray telescope in the energy range between 10$^{11}$ and $10^{15}$ eV.

The first phase of LHAASO will consist of the following major components (see Fig. \ref{fig:lhaaso-layout}):
\begin{itemize}
\item 1.3 km$^2$ array (LHAASO-KM2A) for electromagnetic particle detectors (ED) divided into two parts: a central part including 4931 scintillator detectors 1 m$^2$ each in size (15 m spacing) to cover a circular area with a radius of 575 m and an outer guard-ring instrumented with 311 EDs (30 m spacing) up to a radius of 635 m.
\item An overlapping 1 km$^2$ array of 1146 underground water Cherenkov tanks 36 m$^2$ each in size, with 30 m spacing, for muon detection (MD, total sensitive area $\sim$42,000 m$^2$).
\item A close-packed, surface water Cherenkov detector facility with a total area of about 78,000 m$^2$ (LHAASO-WCDA).
\item 18 wide field-of-view air Cherenkov telescopes (LHAASO-WFCTA).
\end{itemize}

LHAASO is under installation at high altitude (4410 m asl, 600 g/cm$^2$, 29$^{\circ}$ 21' 31'' N, 100$^{\circ}$ 08'15'' E) in the Daochen site, Sichuan province, P.R. China. 
The commissioning of one fourth of the detector will be implemented in 2019 with a sensitivity better than HAWC.
The completion of the installation is expected by the end of 2021.

In Tables~\ref{tab:1-1} and~\ref{tab:1-2} the characteristics of the LHAASO-KM2A array are compared with other experiments. As can be seen, LHAASO will operate with a coverage of $\sim$0.5\% over a 1 km$^2$ area.
The sensitive area of muon detectors is unprecedented and about 17 times larger than CASA-MIA, with a coverage of about 5\% over 1 km$^2$.
By using LHAASO-WCDA as a further muon detector, the total sensitive area for $\mu$-detection will be about 120,000 m$^2$ !

%%%%%%%%% Table 1 %%%%%%%%%%%%%%%%%%%%%%%%%%%%%%%%%%%%%%%%%%%%
\begin{table}[ht!]
\caption{\label{tab:1-1} Characteristics of different EAS-arrays (e.m.)}
\begin{center}
\begin{tabular}{lcccc}
\br
  Experiment &   Altitude& e.m. sensitive & Instrumented & Coverage \\
 &  (m) & area (m$^2$) & area (m$^2$) & \\ \mr
 LHAASO & 4410 & 5.2$\times$10$^3$ & 1.3$\times$10$^6$ & 4$\times$10$^{-3}$ \\
 TIBET AS$\gamma$ & 4300 & 380 & 3.7$\times$10$^4$ & 10$^{-2}$ \\
IceTop & 2835 & 4.2$\times$10$^2$ & 10$^6$ & 4$\times$10$^{-4}$ \\
ARGO-YBJ & 4300 & 6700 & 11,000 & 0.93\\
 &  &  &  & (central carpet)\\
KASCADE & 110 & 5$\times$10$^2$ & 4$\times$10$^4$ & 1.2$\times$10$^{-2}$ \\
KASCADE-Grande & 110 & 370 & 5$\times$10$^5$ & 7$\times$10$^{-4}$ \\
CASA-MIA & 1450 & 1.6$\times$10$^3$ & 2.3$\times$10$^5$ & 7$\times$10$^{-3}$ \\
\br
%\hline
\end{tabular}
\end{center}
\end{table}

\begin{table}[ht!]
\caption{\label{tab:1-2} Characteristics of different EAS-arrays (muons)}
\begin{center}
\begin{tabular}{lcccc}
\br
 Experiment &   Altitude & $\mu$ sensitive & Instrumented& Coverage \\
  &  (m) & area (m$^2$) & area (m$^2$) & \\ \mr
LHAASO & 4410 & 4.2$\times$10$^4$ & 10$^6$ & 4.4$\times$10$^{-2}$ \\
 TIBET AS$\gamma$ & 4300 & 4.5$\times$10$^3$ & 3.7$\times$10$^4$ & 1.2$\times$10$^{-1}$\\
 KASCADE & 110 & 6$\times$10$^2$ & 4$\times$10$^4$ & 1.5$\times$10$^{-2}$ \\
 CASA-MIA & 1450 & 2.5$\times$10$^3$ & 2.3$\times$10$^5$ & 1.1$\times$10$^{-2}$ \\
\br
%\hline
\end{tabular}
\end{center}
\end{table}

%%%%%%%%%%%%%%%%%%%%%%%%%%%%%%%%%%%%%%%%%%%%%%%%%%%%%%%%%

%
%%%%%%%%%%%%%%%%%%%%%%%%%%%%%%%%%%%%%%%%%%%%%%%%%%%%%%%%%%
\begin{figure}[ht!]
  \centerline{\includegraphics[width=0.7\textwidth]{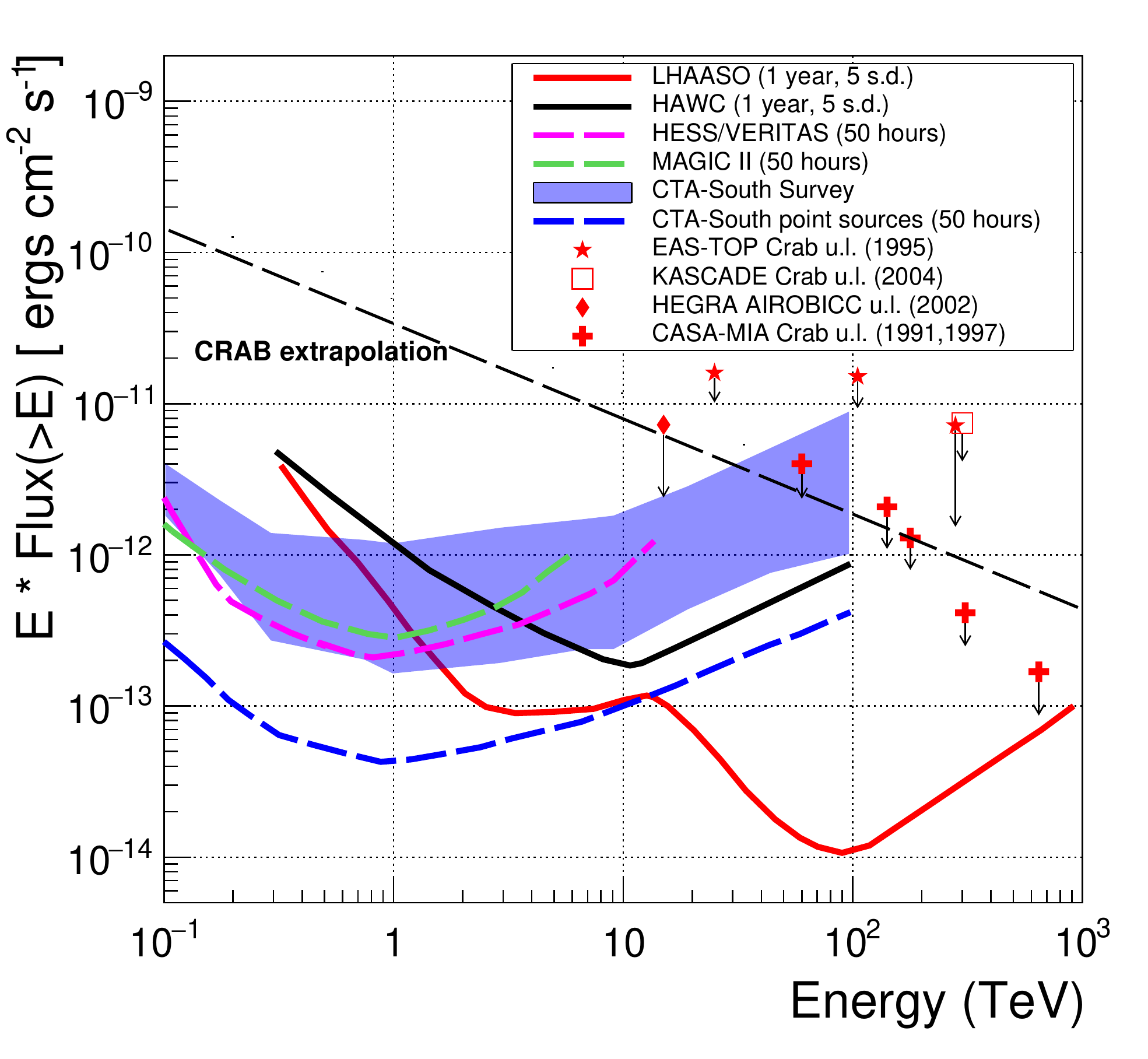} }
\caption{Integral sensitivity of LHAASO as a function of the energy compared to HAWC, HESS, MAGIC II and CTA-South sensitivities. The CTA-South sensitivity to sky survey \cite{cta-survey,hiscore} is also shown.    
Upper limits to ultra high-energy gamma-ray emission set by different experiments in the Northern hemisphere are also reported \cite{eastop97,kascade04,hegra02,casa-mia97}.} 
\label{fig:lhaaso_sens}
\end{figure}
%%%%%%%%%%%%%%%%%%%%%%%%%%%%%%%%%%%%%%%%%%%%%%%%%%%%%%%%%%
%

Fig. \ref{fig:lhaaso_sens} shows the LHAASO sensitivity in detecting a point--like gamma-ray source, compared to that of other experiments. 
The LHAASO sensitivity shows a structure with two minima, reflecting the fact that the observation and identification of photon showers in different energy ranges is carried out by different detectors: the water Cherenkov array (WCDA) in the range $\sim$0.3 -- 10~TeV and the KM2A array above 10 TeV. 
To compare the sensitivities of the experiments reported in Fig. \ref{fig:lhaaso_sens}, it is important to note that, following
the standard convention, the EAS arrays sensitivity is given for a time of one year, while the sensitivity of Cherenkov telescopes is calculated for 50 hours.
As can be seen in the figure, at energies above few TeV, the LHAASO one-year integral sensitivity is better than the sensitivity of MAGIC or HESS in 50 hours. Above 10--20 TeV the LHAASO sensitivity becomes better than that of CTA-South. At 100 TeV, it is about 30 times better than CTA.
Since an IACT can only spend up to $\sim$200 hours per year observing a single source due to solar and lunar constraints, this means that even with the maximum exposure, an IACT still wouldn't be able to match LHAASO sensitivity above a few tens TeV.

It is important to remark that the LHAASO sensitivity shown in the figure also represents the sensitivity for the survey of a large fraction of the sky (56$\%$ of the celestial sphere for observations with a maximum zenith angle of 40$^{\circ}$), while the sensitivity of Cherenkov telescopes only concerns the observation in a region of a few degrees radius that often contains only one source.  
Every day LHAASO (located at latitude 29$^{\circ}$ North) can survey  the declination band from -11$^{\circ}$ to +69$^{\circ}$
that includes the galactic plane in the longitude interval from +20$^{\circ}$ to +225$^{\circ}$.
HAWC has a sensitivity $\sim$4 times lower than that expected for LHAASO in the 1--10 TeV region, but more than 100 times lower at 100 TeV.
Concerning Cherenkov telescopes, their limited FoV and duty cycle prevent a survey of large regions of the sky.  
To compare the effective sensitivities in sky survey, one has to take into account the fact that a Cherenkov telescope must scan the whole region under study with different pointings. The number of pointings determines the maximum observation time that can be dedicated to any source. 
If we consider a survey of the Galactic Plane in a galactic longitude interval of $\sim$200$^{\circ}$, a reasonable number of pointing is $\sim$100.  Assuming a total observation time of $\sim$1300 hours/year, a full year dedicated to the survey allows an exposure of $\sim$13 hours for source.
This time is reduced to less than 2 hours for an {\it all sky survey} of $\sim\pi$ sr of solid angle, that requires approximately $\sim$800 pointings.
The reduced observation time causes an increase of the minimum detectable flux, as shown in the blue band of Fig.\ref{fig:lhaaso_sens}, where the lower limit of the band refers to a Galactic plane survey and the upper limit to an {\it all sky survey} of $\pi$ sr \cite{cta-survey,hiscore}.

%%%%%%%%% Table 2 %%%%%%%%%%%%%%%%%%%%%%%%%%%%%%%%%%%%%%%%%%%%
\begin{table}[ht!]
\caption{\label{tab:2} Performance comparison between LHAASO-KM2A and CASA-MIA experiments.}
\begin{center}
\begin{tabular}{lll}
\br
Experiment &  LHAASO-KM2A & CASA-MIA \\ \mr
Angular resolution & 0.3$^{\circ}$ (100 TeV) & 2$^{\circ}$ (100 TeV) \\
 & 0.2$^{\circ}$ (1 PeV) & $\sim$0.5$^{\circ}$ (646 TeV) \\
$\mu$ detector sensitive area (m$^2$) & 42,000    & 2500 \\
EAS array instrumented area (m$^2$)  & 1.3$\times$10$^6$ & 230,000 \\
$\mu$ detector coverage & 4.4$\times$10$^{-2}$ & 1.1$\times$10$^{-2}$  \\
Background hadron & $\sim$10$^{-5}$ ($\geq$100 TeV) & 10$^{-2}$ (178 TeV) \\
surviving efficiency &   & 2$\times$10$^{-4}$ (646 TeV) \\
\br
\end{tabular}
\end{center}
\end{table}
%%%%%%%%%%%%%%%%%%%%%%%%%%%%%%%%%%%%%%%%%%%%%%%%%%%%%%%%%

In Fig.~\ref{fig:lhaaso_sens} the upper limits set by different experiments to high energy gamma-ray emission in the Northern hemisphere are reported \cite{eastop97,kascade04,hegra02,casa-mia97}.
In five years of observations, the CASA-MIA experiment sets the lowest upper limits to the flux from the Crab Nebula around and above 100 TeV \cite{casa-mia97}. Beyond 1 PeV, the IceTop/IceCube experiments, located at the South Pole, reports a minimum observable gamma ray flux ranging from $\sim$10$^{-19}$ to 10$^{-17}$ photons s$^{-1}$ cm$^{-2}$ TeV$^{-1}$ (depending on the source declination) for sources on the galactic plane in 5 years of measurements \cite{icecube13}.
Table~\ref{tab:2} compares the performance of the LHAASO-KM2A array with the CASA-MIA experiment. 
At 100 TeV, the angular resolution of the LHAASO-KM2A array for gamma rays is $\sim$7 times better than that of CASA-MIA, and the area is $\sim$4 times larger. The efficiency in background rejection is about 2$\times$10$^3$ times better in LHAASO, due to the larger muon detector area.
According to expression $F_{min} \propto \frac{ \sigma}{ Q \times \sqrt{A \times T}}$, the LHAASO sensitivity is $\sim$500 times better than that of CASA-MIA at 100 TeV.
With this sensitivity, LHAASO can perform measurements of the high energy tails of emission spectra for the majority of the known TeV galactic sources visible from its location with unprecedented sensitivity.

LHAASO will enable studies in CR physics and gamma-ray astronomy that are unattainable with the current suite of instruments:
\begin{itemize}
\item[1)] LHAASO will perform an \emph{unbiased sky survey of the Northern sky} with a detection threshold better than 10\% Crab units at sub-TeV/TeV and 100 TeV energies in one year. This unique detector will be capable of continuously surveying the $\gamma$-ray sky for steady and transient sources from a few hundred GeV to the PeV energy domain.
From its location LHAASO will observe at TeV energies and with high sensitivity about 30 of the sources catalogued by Fermi-LAT at lower energy, monitoring the variability of 15 AGNs (mainly blazars) at least.
\item[2)] The sub-TeV/TeV LHAASO sensitivity will allow to observe AGN flares that are unobservable by other instruments, including the so-called TeV orphan flares. 
\item[3)] LHAASO will study in detail the high energy tail of the spectra of most of the $\gamma$-ray sources observed at TeV energies, opening for the first time the 100--1000 TeV range to the direct observations of the high energy cosmic ray sources.
\item[4)] LHAASO will map the Galactic \emph{diffuse gamma-ray emission} above few hundred GeV and thereby measure the CR flux and spectrum throughout the Galaxy with high sensitivity. 
The measurement of the space distribution of diffuse $\gamma$-rays will allow to trace the location of the CR sources and the distribution of interstellar gas.
\item[5)] The high background rejection capability in the 10 -- 100 TeV range will allow LHAASO to measure the \emph{isotropic diffuse flux of ultrahigh energy $\gamma$ radiation} expected from a variety of sources including Dark Matter and the interaction  of 10$^{20}$ eV CRs with the 2.7 K microwave background radiation. 
In addition, LHAASO will be able to achieve a limit below the level of the IceCube diffuse neutrino flux at 10 -- 100 TeV, thus constraining the origin of the IceCube astrophysical neutrinos.
\item[6)] LHAASO will allow the reconstruction of the energy spectra of different CR mass groups in the 10$^{12}$ -- 10$^{17}$ eV with unprecedented statistics and resolution, thus tracing the light and heavy components through the knee of the all-particle spectrum.
\item[7)] LHAASO will allow the measurement, for the first time, of the CR anisotropy across the knee separately for light and heavy primary masses.
\item[8)] The different observables (electronic, muonic and Cherenkov components) that will be measured in LHAASO will allow a detailed investigation of the role of the hadronic interaction models, therefore investigating if the EAS development is correctly described by the current simulation codes.
\item[9)] LHAASO will look for signatures of WIMPs as candidate particles for DM with high sensitivity for particles masses above 10 TeV. Moreover, axion-like particle searches are planned, where conversion of gamma-rays to/from axion-like particles can create distinctive features in the spectra of gamma-ray sources and/or increase transparency of the universe by reducing the Extragalactic Background Light (EBL) absorption. 
Testing of Lorentz invariance violation as well as the search for Primordial Black Holes and Q--balls will also be part of the scientific programme of the experiment.
\end{itemize} 

In the next decade CTA-North and LHAASO are expected to be the most sensitive instruments to study Gamma-Ray Astronomy in the Northern hemisphere from about 20 GeV up to PeV.

\subsection{TAIGA-HiSCORE experiment}

The new TAIGA-HiSCORE non-imaging Cherenkov array aims to detect air showers induced by gamma rays above 30 TeV and to study cosmic rays above 100 TeV. 
TAIGA-HiSCORE is made of integrating air Cherenkov detector stations with a wide FoV ($\sim$0.6 sr), placed at a distance of about 100 m to cover a final area of $\sim$5 km$^2$.
%
%%%%%%%%%%%%%%%%%%%%%%%%%%%%%%%%%%%%%%%%%%%%%%%%%%%%%%%%%%
\begin{figure}[ht!]
\centerline{\includegraphics[width=0.8\textwidth,clip]{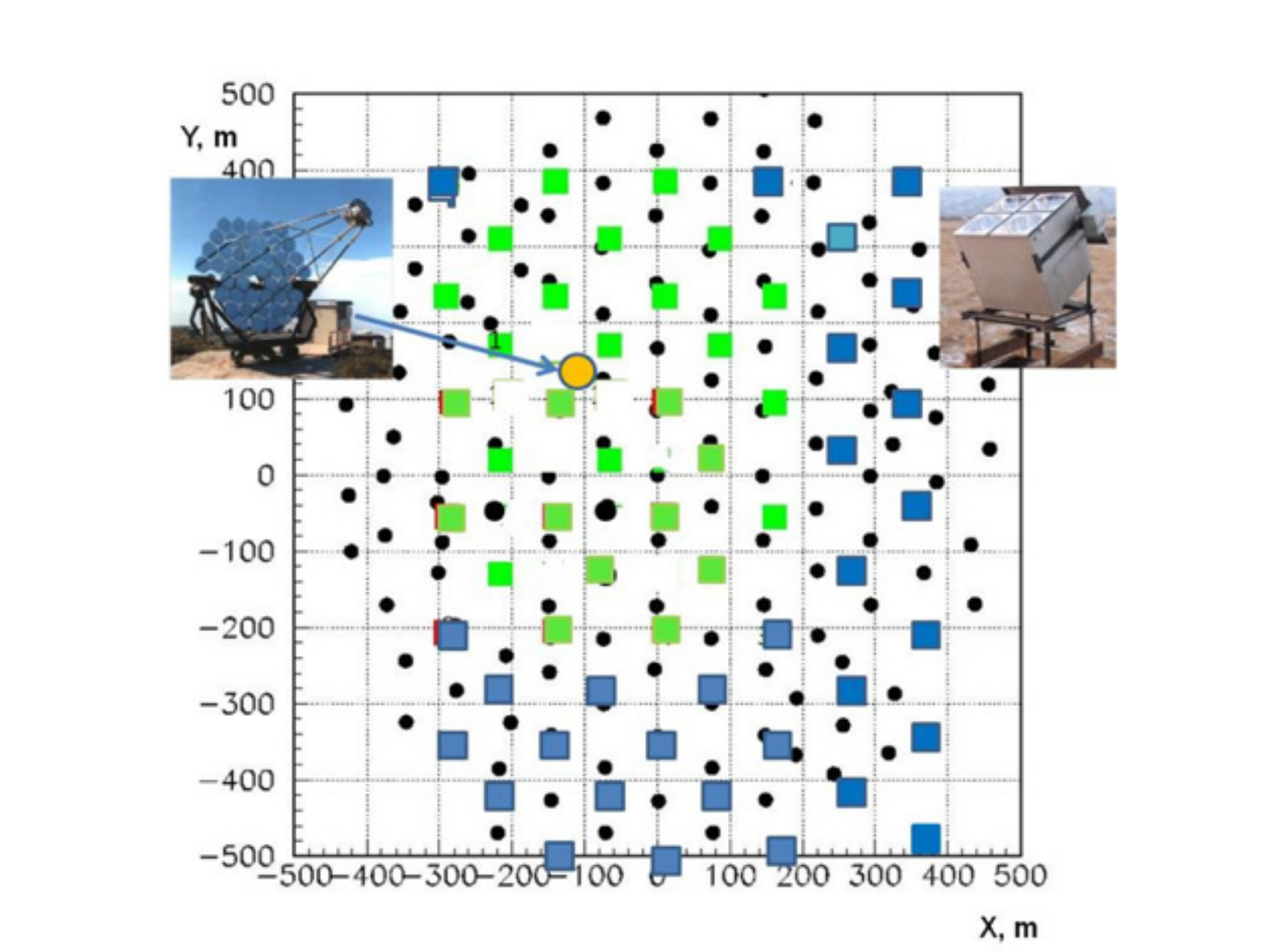} }
\caption{The TAIGA-prototype-2017. Green squares: optical stations of TAIGA-HiSCORE installed in 2014. Blue squares: optical stations to be installed in 2017. Yellow circle: position of the first TAIGA-IACT. Small black circles: optical detectors of Tunka-133.} 
\label{fig:hiscore}       % Give a unique label
\end{figure}
%%%%%%%%%%%%%%%%%%%%%%%%%%%%%%%%%%%%%%%%%%%%%%%%%%%%%%%%%%
%

The TAIGA-HiSCORE array is part of the gamma-ray observatory TAIGA (Tunka Advanced Instrument for cosmic ray physics and Gamma Astronomy). TAIGA is currently under construction in the Tunka valley, about 50 km from Lake Baikal in Siberia, Russia \cite{taiga16}. The key advantage of the TAIGA will be the hybrid detection of EAS Cherenkov radiation by the wide-angle detector stations of the TAIGA-HiSCORE array and by the Imaging Air Cherenkov Telescopes of the TAIGA-IACT array. TAIGA comprises also the Tunka-133 array and will furthermore host up a net of surface and underground stations for measuring the muon component of air showers.
The principle of the TAIGA-HiSCORE detector is the following: the detector stations measure the light amplitudes and full time development of the air shower ligth front up to distances of several hundred meters from the shower core.

Currently TAIGA-HiSCORE array is composed of more than 40 detector stations distributed in a regular grid over a surface area of $\sim$0.5 km$^2$ with an inter-station spacing of about 106 m (prototype array, see Fig. \ref{fig:hiscore}).
Each optical station contains four large area photomultipliers with 20 or 25 cm diameter. Each PMT has a light collector Winston cone with 0.4 m diameter and 30$^{\circ}$ viewing angle (FoV of $\sim$0.6 sr). Plexiglass is used on top to protect the PMTs against dust and humidity. A total station light collection area is 0.5 m$^2$ \cite{hiscore}.

Before the winter season 2017--2018 the TAIGA configuration will include 60 wide angle stations arranged over an area of 0.6 km$^2$, and one single IACT. The expected integral sensitivity for 200 hours of a source observation (about 2 seasons of operation) in the range 30--200 TeV is about 10--12 erg cm$^2$ sec$^{-1}$ \cite{taiga17}.

\section{What's Next ?}

All the survey instruments mentioned in the previous sections are located in the Northern hemisphere.
The construction of a new wide FoV detector at sufficiently Southern latitude to continuously monitor the Galactic Center and the Inner Galaxy should be a high priority \cite{sgso,disciascio-ricap18}. 

A new Southern Hemisphere wide FoV detector, to be fully complementary to CTA-South and to carry out an integrated study of gamma and nuclei induced showers, needs 
\begin{enumerate}
\item an energy threshold of $\sim$100 GeV, to be a transient factory;
\item a sensitivity at few percent Crab flux level below the TeV, to have high exposure for flaring activity;
\item an angular resolution of $\sim$1$^{\circ}$ at the threshold, to reduce source confusion in the Inner Galaxy;
\item to survey the gamma sky at 100 TeV with a capability to discriminate the background at a level of 10$^{-5}$ to observe the knee in the energy spectrum of the gamma diffuse emission at $\approx$300 TeV, corresponding to the knee in the CR all-particle spectrum, in different regions of the Galactic Plane;
\item to be able to discriminate different primary masses in the knee energy range to measure the proton knee and investigate the maximum energy of accelerated particles in CR sources and to observe the CR anisotropy as a function of the particle rigidity.
\end{enumerate}
Is this possible ?\\
As discussed in Section 4, the main parameters to push down the sensitivity to gamma-ray sources are: (1) the energy threshold; (2) the angular resolution; (3) the gamma/hadron relative trigger efficiency; (4) the effective area for photon detection; (5) the background rejection capability.

Different groups are proposing different ideas for a new wide FoV in the South, ALTO, ALPACA, LATTES and STACEX \cite{alto,alpaca,mostafa,lattes}.

ALTO is the name given by Linnaeus University to a project to build a wide-field gamma-ray observatory at high-altitude in the southern hemisphere of the same technological "family" of HAWC \cite{alto,alto2}.
For ALTO, there are a number of points through which the current design of HAWC can be improved as: the altitude of the observatory, the use of a layer of scintillator below the water tank, the construction of smaller tanks, and the use of more precise electronics and time-stamping.
The construction of smaller tanks with respect to HAWC will allow to have a finer-grained view of the shower particles on the ground which helps in the reconstruction of the arrival direction of the incoming event and in the background rejection. The current design of the ALTO water tank is hexagonal, so as to be close-packed.
A layer of scintillator below the water tank is very important in order to be able to "tag" the passage of muons, which are the almost unambiguous signature of the nature of the particle cascade, as background proton-initiated showers are muon-rich. The implementation of this new muon-tagging" allows to reach an increased signal over background discrimination by analysis, and thus allows an increase of the sensitivity of the detection technique. 

The ALPACA experiment is a new project aimed at the observation of cosmic rays and gamma rays, launched between Bolivia and Japan in 2016 \cite{alpaca,alpaca2}. They are planning to construct an 83,000 m$^2$ surface air-shower array and a 5,400 m$^2$ underground muon detector array, on a highland at the altitude of 4,740 m halfway up Mt. Chacaltaya on the outskirts of La Paz, Bolivia. The muon detector array enables us to select muon-poor air showers (i.e. air showers induced by primary gamma rays) and thus improve the gamma-ray sensitivity of the air-shower array. The project is named ALPACA (Andes Large-area PArticle detector for Cosmic-ray physics and Astronomy), after the animal that inhabits South America. The layout of the array is similar to the Tibet AS$\gamma$ experiment, in operation in Tibet with increasing areas since 1990.

An interesting proposal is to built a hybrid detector to exploit the characteristics of two detectors, RPCs and Water Cherenkov Detectors (WCD).
The LATTES project will consist of one layer of RPCs on top of WCD of small dimensions \cite{lattes}.
%
%%%%%%%%%%%%%%%%%%%%%%%%%%%%%%%%%%%%%%%%%%%%%%%%%%%%%%%%%%
\begin{figure}[ht!]
\centerline{\includegraphics[width=0.8\textwidth,clip]{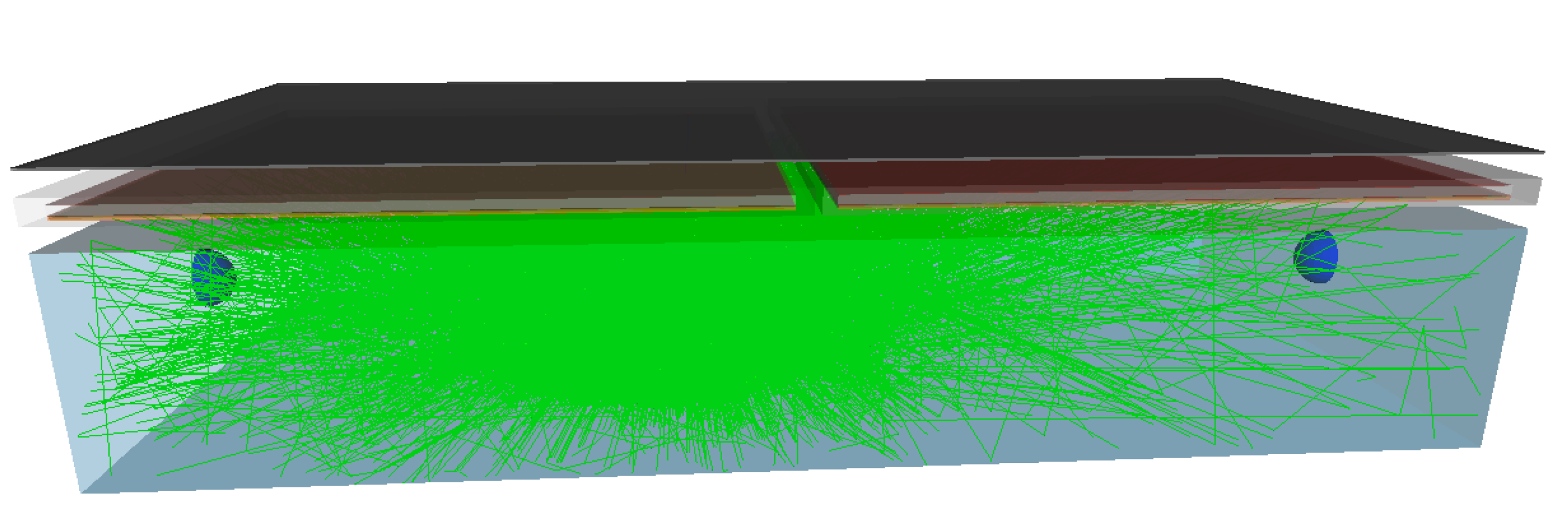} }
\caption{Basic LATTES detector station, with one WCD covered with RPCs and a thin slab of lead \cite{lattes}. 
The green lines show the tracks of the Cherenkov photons inside the water.} 
\label{fig:lattes-basic_station}       % Give a unique label
\end{figure}
%%%%%%%%%%%%%%%%%%%%%%%%%%%%%%%%%%%%%%%%%%%%%%%%%%%%%%%%%%
%

The basic element, shown in Fig. \ref{fig:lattes-basic_station}, is composed of one WCD, with a rectangular horizontal surface of 
3 m$\times$1.5 m and a depth of 0.5 m, covered by two RPCs, each with a surface of  (1.5$\times$1.5) m$^2$, with a lead layer on top (5.6 mm) to exploit the secondary photons conversion.
The proposed RPCs are based on glass, different from the bakelite RPCs operated for more than 10 years at 4300 m asl by the ARGO-YBJ collaboration \cite{disciascio-rev} and extensively used at LHC.

The proposed RPCs are designed to work at low gas flux, (1--4) cc/min, at harsh outdoor environment, and demanding very low maintenance services. Their intrinsic time resolution was measured to be better than 1 ns.
The WCD read-out will be provided by two 15 cm PMTs at both ends of the smallest vertical face of the tank.
The read-out of the RPCs will be provided by 16 charge collecting pads.
This hybrid detector is expected to improve the trigger selection at low energies and the rejection of the background of charged nuclei.
The shower energy will be reconstructed from the total signal, defined as the sum of the number of photoelectrons in all WCD stations.
The proposed experiment will consist of an array of 60$\times$30 stations, covering an effective area of about 10,000 m$^2$ located at 5200 m asl.
The different detectors will be separated by a small distance (roughly 0.5 m) to allow access to PMTs and RPCs. 
An angular resolution better than 2$^{\circ}$ is expected in the 100 GeV range.
A 1-year sensitivity at level of 15\% of the Crab Nebula flux is expected in the 100 -- 400 GeV energy range.

The key to lower the energy threshold is to locate the detector at extreme altitude (about 5000 m asl for a threshold in the 100 GeV range). But the energy threshold, as well as the angular resolution, depends also on the coverage (the ratio between the detection area and the instrumented one), on the granularity of the read-out, on the particular type of detector and on the trigger logic.

The ARGO-YBJ experiment, combining the full coverage approach at high altitude with a high granularity of the read-out (about 15,000 strips 7$\times$62 cm$^2$ wide), sampled 100 GeV $\gamma$-induced showers with an efficiency of about 70\% at 4300 m a.s.l. . 
The median energy of the first multiplicity bin (20-40 fired pads) for photons with a Crab-like energy spectrum was 340 GeV \cite{argo-crab}.
The granularity of the read-out at cm level allowed to sample events with only 20 fired pads, out of 15,000, with a background-free topological-based trigger logic.
A water Cherenkov based facility will hardly be able to lower the energy threshold below $\sim$500 GeV. 

Therefore, a full coverage approach based on the RPC technology is one of the most interesting solution for a new survey instrument in the South. A hybrid detector with RPCs coupled to a HAWC/LHAASO - like water Cherenkov detector, will allow to lower the energy threshold at 100 GeV level, to measure the arrival direction with two independent techniques, to exploit the HAWC/LHAASO approach to reject the background, and to study also CR physics with nuclei induced events up to 10 PeV. 

The ARGO-like RPCs should be an important element of a future experiment in the South, possibly coupled to HAWC/LHAASO - like water Cherenkov detectors to exploit the added values of two experimental approaches. This is the aim of the STACEX proposal.

Science case and experimental solutions for a survey instrument in the South are under discussion in the framework of the Southern Gamma-ray Survey Observatory (SGSO) alliance \cite{sgso}. Preliminary calculations of the sensitivity of a water Cherenkov based array is shown in Fig. \ref{fig:sensitivity}.

Science case and experimental solutions for a survey instrument in the South are under discussion in the framework of the Southern Gamma-ray Survey Observatory (SGSO) alliance. Science case and preliminary calculations of the sensitivity of a water Cherenkov based array, covering an area of 221,000 m$^2$ with a fill-factor of 8\% and located at 5000 m asl, are presented in \cite{sgso} (see Fig. \ref{fig:sensitivity}).

%\section*{Acknowledgements}
\ack

I would like to thank the organizers of the \emph{ISAPP-Baikal Summer School "Exploring the Universe through multiple messengers"} for their invitation and the warm hospitality in Bol'shie Koty.

\section*{References}
 
%\bibliography{refs}

%\end{document}

\expandafter\ifx\csname url\endcsname\relax
  \def\url#1{{\tt #1}}\fi
\expandafter\ifx\csname urlprefix\endcsname\relax\def\urlprefix{URL }\fi
\providecommand{\eprint}[2][]{\url{#2}}

\end{document}